\documentclass[aps,twocolumn,pra,superscriptaddress,amsmath,amssymb]{revtex4-1} 
\usepackage{dcolumn}
\usepackage{bm}
\usepackage{amsmath}
\usepackage{txfonts}
\usepackage[T1]{fontenc}
\usepackage{xspace}
\usepackage{ulem}
\usepackage{comment}
\usepackage{bbold}
\newcommand{\means}[1]{\langle#1\rangle}

\ifx\pdfoutput\undefined
\usepackage[dvipdfmx]{graphicx}
\usepackage[dvipdfmx]{hyperref}
\usepackage[dvipdfmx]{color}
\usepackage[dvipdfmx]{xcolor}
\else
\usepackage{graphicx}
\usepackage{hyperref}
\usepackage{color}
\usepackage{xcolor}
\fi
\begin{document}
\let\emph\textit

\title{
Spin dynamics in the Kitaev model with disorder: Quantum Monte Carlo study of dynamical spin structure factor, magnetic susceptibility, and NMR relaxation rate
  }
\author{Joji Nasu}
\affiliation{
  Department of Physics, Yokohama National University, Hodogaya, Yokohama 240-8501, Japan
}
\affiliation{
  PRESTO, Japan Science and Technology Agency, Honcho Kawaguchi, Saitama 332-0012, Japan
}
\author{Yukitoshi Motome}
\affiliation{
  Department of Applied Physics, University of Tokyo,
  Bunkyo, Tokyo 113-8656, Japan
}

\date{\today}
\begin{abstract}
  We investigate the impact of two types of disorder, bond randomness and site dilution, on the spin dynamics in the Kitaev model on a honeycomb lattice.
  The ground state of this model is a canonical quantum spin liquid with spin fractionalization into two types of quasiparticles, itinerant Majorana fermions and localized fluxes, for which the spin dynamics provides a good probe of the fractionalization.
  Using unbiased quantum Monte Carlo simulations, we calculate the temperature evolution of the dynamical spin structure factor, the magnetic susceptibility, and the NMR relaxation rate while changing the strength of disorder systematically.
  In the dynamical spin structure factor, we find that the two types of disorder affect seriously the low-energy peak dominantly originating from the flux excitations, rather than the high-energy continuum from the Majorana excitations, in a different way:
  The bond randomness softens the peak to the lower energy with broadening, which suggests the closing of the spin gap, whereas the site dilution smears the peak and in addition develops the other sharp peaks inside the spin gap including the zero energy. 
  We show that the zero-energy spin excitations, which originate from the Majorana zero modes induced around the site vacancies, survive up to the temperature comparable to the energy scale of the Kitaev interaction.
  We also find that the two types of disorder affect the low-temperature behavior of the magnetic susceptibility and the NMR relaxation rate in a different way.
  For the bond randomness, the low-temperature susceptibility does not show any qualitative change against the weak disorder, but it changes to divergent behavior while increasing the strength of disorder.
  We find that this crossover corresponds to the softening of the low-energy peak in the dynamical structure factor.
  Similar distinct behaviors for the weak and strong disorder are observed also in the NMR relaxation rate;
  an exponential decay changes into a power-law decay.
  In contrast, for the site dilution, we find no such crossover; divergent behavior in the susceptibility and a power-law decay in the NMR relaxation rate appear immediately with the introduction of the site dilution, which is also attributed to the emergence of the Majorana zero modes.
  We discuss the relevance of our results to experiments for the Kitaev candidate materials with disorders.
  The peculiar magnetic responses found by the present systematic analysis would be helpful to not only identify the dominant type of disorder in real materials but also examine the experimental realization of the Kitaev spin liquid by introducing disorder. 
\end{abstract}
\maketitle

\section{Introduction}
\label{sec:Introduction}

Quantum many-body effects bring about peculiar phenomena unpredictable from free particles, not only in the ground state but also in the excited states.
A representative phenomenon in condensed matter physics is fractionalization of elementary particles, electrons, in strongly correlated systems, appearing as elementary excitations from a nontrivial ground state.
The celebrated example is found in the fractional quantum Hall effect in a two-dimensional electron system under a strong magnetic field, where the ground state is topologically nontrivial and the excitations are described by composite quasiparticles with a fractional charge of electron~\cite{moore1991nonabelions,Stormer1999,Read2000,jain2015composite}.
Another example is a quantum spin liquid (QSL) in insulating magnets, which is a highly quantum entangled state achieved when any symmetry of the system is not broken down to zero temperature ($T$) due to strong quantum fluctuations~\cite{Anderson1973153,fazekas1974ground,ISI:000275366100033,Savary2017}.
In the QSL, the spin excitations are anticipated to be fractionalized into, e.g., spinons, which carry spin-$1/2$ each but no charge, and visons, which carry neither spin nor charge~\cite{Read1989,Read1991,Wen1991,Baskaran1988,Senthil2000,PhysRevLett.96.060601}.
Of particular interest is the realizations in higher dimensions more than one, where the spinons may form the Fermi surfaces~\cite{Dagotto1988,Wen1996,Hermele2004}
.
This is expected to be observed as the $T$-linear specific heat, despite the insulating state.
Also, the fractionalization is expected to give rise to unconventional continuum spectrum in the dynamical spin structure factor, instead of the coherent spin-wave excitations~\cite{kohno2007spinons,Thielemann2009,Klauser2011}.
Such behaviors are believed to be a hallmark of QSLs in experiments~\cite{ISI:000278318600025,han2012fractionalized,PhysRevLett.112.177201,watanabe2016emergence}.
On the other hand, disorders, which inevitably exist in real materials, may also lead to magnetically disordered ground states and unconventional low-energy excitations.
Therefore, to capture the inherent nature of the QSL, one needs to distinguish the effects of intrinsic quantum fluctuations and extrinsic disorders.
This is, however, a big challenge in both experiment and theory; it is hard to find a good candidate with controlled less disorders in experiments, and it is also theoretically difficult to perform the calculations with sufficient accuracy to extract the intrinsic effects.

A good platform to address this issue is found in the recent research for the Kitaev honeycomb model and the candidate materials~\cite{RevModPhys.87.1,Trebst2017pre,Hermanns2018rev,Knolle2019rev,takagi2019rev,Motome2020rev}. 
The Kitaev model is defined by Ising-type interactions between neighboring $S=1/2$ spins dependent on the bond directions, whose ground state is exactly obtained as a QSL in two dimensions~\cite{Kitaev2006}.
The elementary excitations from the QSL ground state are described by two types of quasiparticles emergent from spin fractionalization: itinerant Majorana fermions and localized fluxes, corresponding to spinons and visons, respectively.
The characteristic Kitaev-type interaction is thought to be realized in transition metal compounds with strong spin-orbit coupling, through the exchange processes via the ligands on the edge-sharing network of octahedra~\cite{PhysRevLett.102.017205}.
The iridium oxides $A_2$IrO$_3$ ($A=$Li, Na)~\cite{PhysRevLett.105.027204,PhysRevB.82.064412,PhysRevLett.108.127203,PhysRevB.88.035107,PhysRevLett.110.097204,1367-2630-16-1-013056,PhysRevLett.113.107201,Winter2016} and the ruthenium compound $\alpha$-RuCl$_3$~\cite{PhysRevB.90.041112,PhysRevB.91.094422,PhysRevLett.114.147201,Johnson2015,PhysRevB.91.144420,Cao20016,yadav2016kitaev,Winter2016,PhysRevB.93.155143,Koitzsch2016} have been intensively investigated as the representatives of Kitaev candidate materials.
While these materials do not realize the QSL ground state and exhibit magnetic orders at low $T$, the thermodynamic properties and spin dynamics show good agreement with the theoretical results for the Kitaev model~\cite{PhysRevB.92.115122,PhysRevLett.112.207203,PhysRevB.92.115127,PhysRevLett.113.187201,winter2017breakdown,Song2016,Nasu2016nphys,Halasz2016,yoshitake2016,Nasu2017,Yoshitake2017PRBb,Yoshitake2017PRBa,Udagawa2018,Yoshitake2020,Kasahara2018,kasahara2018majorana,Yokoi2020pre}.
This suggests that the other interactions besides the Kitaev type, which induce the magnetic orders, are relatively small and the fractionalization into the Majorana fermions and fluxes indeed occurs under the dominant Kitaev-type interaction.

Extrinsic disorders have been intentionally introduced to these candidate materials as an attempt to suppress the magnetic orderings due to the subsidiary interactions and stabilize the Kitaev QSL.
For example, chemical substitutions of the magnetic Ru$^{3+}$ ions have been investigated for $\alpha$-RuCl$_3$.
For the replacement by nonmagnetic Ir$^{3+}$ ions, the magnetic order is suppressed~\cite{Lampen-Kelley2017} and the power-law behavior is observed in the thermodynamic quantities at low $T$, suggesting a weakly-divergent low-energy excitations induced by the substitution~\cite{Do2018,Do2020}.
In the case of the replacement by the other magnetic Cr$^{3+}$ ions, spin-glass like behavior was reported~\cite{Bastien2019}.
Similar suppression of the magnetic order has also been observed in the iridium oxides.
In the solid solutions (Na$_{1-x}$Li$_x$)$_2$IrO$_3$, the magnetic transition temperature is lowered in the intermediate $x$ region~\cite{Cao2013,Manni2014,Rolfs2015,gupta2016raman,Hermann2017,Simutis2018}. 
Moreover, the replacements of the Ir$^{4+}$ ions by Ru$^{4+}$ and Ti$^{4+}$ in $A_2$IrO$_3$ ($A=$Li, Na) were shown to suppress the magnetic order and eventually realize a spin glass or dimerized ground state~\cite{Lei2014,Manni2014a}.
Another iridium oxide H$_3$LiIr$_2$O$_6$, which does not show any magnetic order down to the lowest $T$~\cite{Kitagawa2018nature}, is discussed to have randomness due to disorder in the hydrogen positions~\cite{Yadav2018,LiYing2018,wang2020possible}.
This material shows a peculiar asymptotic behavior of the specific heat at low $T$, which changes in an external magnetic field.
All of these results suggest that the disorder not only suppresses the magnetic ordering but also can yield intriguing phenomena in the Kitaev magnets.

The above disorders are roughly categorized into two types, bond randomness and site randomness.
The disorders in (Na$_{1-x}$Li$_x$)$_2$IrO$_3$ and H$_3$LiIr$_2$O$_6$ can be classified to the former, while the others by the substitutions of the magnetic ions are to the latter.
To understand these disorder effects, the Kitaev model with randomness has been investigated theoretically.
For instance, the bond randomness was shown to affect the low-energy part of the dynamical spin correlator at $T=0$~\cite{Zschocke2015} and give rise to the low-energy divergence in the density of states (DOS) of Majorana fermions, which leads to the power-law behavior of the specific heat at low $T$~\cite{Knolle2019disorder}.
The site dilution also raises peculiar behavior in the thermodynamic quantities, such as the power-law $T$ dependence of the specific heat~\cite{Kao2020}, the logarithmic divergence of the magnetic susceptibility~\cite{Willans2010,Willans2011,Santhosh2012}, and emergence of the spin glass behavior~\cite{Andrade2014,Andrade2020}.
An intriguing point specific to the site dilution is that unpaired Majorana zero modes are induced around the vacancy sites~\cite{Petrova2013,Petrova2014,brennan2016lattice}, which were suggested to be observed in the dynamical spin correlations~\cite{Udagawa2018,Otten2019}.
Such different effects by the bond randomness and the site dilution were recently studied by the authors in a systematic way for the specific heat and the thermal transport~\cite{Nasu2020disorder}.
While a lot of theoretical studies were devoted to the disorder effects on the Kitaev QSL, the evolution of the spin dynamics against the disorder has not been systematically investigated thus far for the two types of disorder, despite the importance for understanding the intrinsic and extrinsic nature of the quantum disordered states in the candidate materials.

In this paper, we investigate the disorder effect on the spin dynamics in the Kitaev QSL, with an emphasis on the different aspects of the two types of disorder, the bond randomness and the site dilution, in the $T$ dependences of the experimental observables. 
Specifically, by employing a quantum Monte Carlo method based on the Majorana fermion representation, where both intrinsic quantum fluctuations and extrinsic disorder effects are fully taken into account, we calculate the dynamical spin structure factor, the magnetic susceptibility, and the nuclear magnetic resonance (NMR) relaxation rate.
For the dynamical spin structure factor, we find that the low-energy spectra show clearly different responses to the two types of disorder. 
In the case of the bond randomness, the low-energy peak, which originates from the gapped flux excitations in the pristine case, is broadened and shifted to the lower-energy side by increasing the strength of the randomness.
On the other hand, in the case of the site dilution, the peak is smeared with no apparent energy shift, but the other sharp peak is developed at zero energy associated with the appearance of the Majorana zero modes.
We find that the newly-developed elastic peak survives up to the temperature comparable to the energy scale of the Kitaev interaction.
We also find contrasting behaviors against the two types of disorder in the magnetic susceptibility and the NMR relaxation rate.
While the low-$T$ susceptibility shows the Van Vleck-type $T$ dependence for the weak bond randomness in the calculated $T$ range, it rapidly changes into a divergent behavior while increasing the randomness.
Correspondingly, the low-$T$ behavior of the NMR relaxation rate also shows a substantial change from an exponential decay to a power-law one.
We show that these crossovers are closely related with the softening of the low-energy peak in the dynamical spin structure factor.
On the other hand, in the case of the site dilution, the divergent behavior in the susceptibility and the power-law decay of the NMR relaxation rate appear immediately with the introduction of the disorder, which is also ascribed to the appearance of the Majorana zero modes associated with the vacancies.
Carefully analyzing the low-$T$ data, we show that the asymptotic forms of these quantities can be fitted by peculiar exponents.
Our results would be helpful to discuss the intrinsic and extrinsic nature in the Kitaev candidate materials with disorders.

This paper is organized as follows.
In the next section, we introduce the model and its fundamental properties.
In Sec.~\ref{subsec:model_Majorana}, we describe a mapping of the Kitaev quantum spin model onto a Majorana fermion system coupled with local variables.
The two types of disorder addressed in the present study are introduced in Sec.~\ref{subsec:model_disorder}.
In Secs.~\ref{sec:Signatures-spin-fractionalization} and \ref{sec:thermodynamics-transport}, we briefly review the finite-$T$ properties of the pristine Kitaev model and the disorder effects on the thermodynamics and thermal transport, respectively.
The method of the numerical calculations is described in Sec.~\ref{sec:method}. 
The framework of the Monte Carlo (MC) simulation is given in Sec.~\ref{sec:MC-simulations}.
In Sec.~\ref{sec:dynamical-spin-correlation}, we describe the way of evaluating the dynamical spin correlations in the MC simulations.
We remark a special care in the calculations for the site dilution in Sec.~\ref{sebsec:unpaired}.
The formulas for the dynamical spin structure factor, the magnetic susceptibility, and the NMR relaxation rate are derived in Secs.\ref{sec:dynamical-spin-structure-factor}, ~\ref{sec:Magnetic-susceptibility}, and \ref{sec:NMR-relaxation-rate}, respectively.
The results are given in Sec.~\ref{sec:result}.
We present the dependences on the strength of the bond randomness and the density of vacancies of the dynamical spin structure factor, the magnetic susceptibility, and the NMR relaxation rate as function of temperature in Secs.~\ref{sec:dsf}, \ref{sec:result-suscep}, and \ref{sec:NMR_result}, respectively.
In Sec~\ref{sec:discussion}, we discuss the relevance to experimental results. Finally, Sec.~\ref{sec:summary} is devoted to the summary.

\section{Model}

\subsection{Hamiltonian and Majorana representation}
\label{subsec:model_Majorana}

\begin{figure}[t]
  \begin{center}
  \includegraphics[width=\columnwidth,clip]{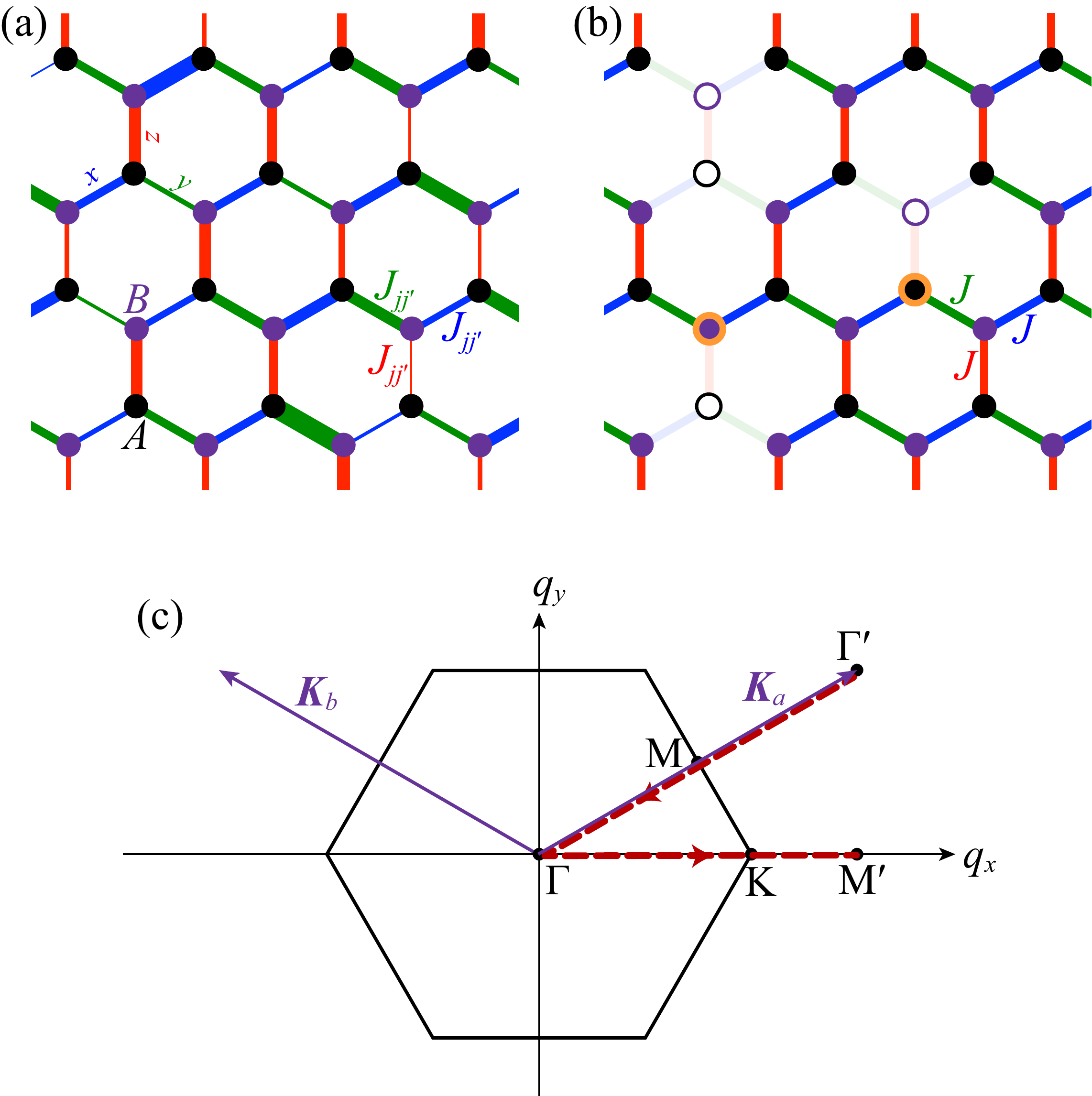}
  \caption{
Schematic pictures of the Kitaev model in Eq.~\eqref{eq:HamilS} with (a) the bond randomness and (b) the site dilution.
The $A$ and $B$ sublattice sites are represented by the black and purple circles, respectively, and the three kinds of the bonds, $x$, $y$, and $z$, are depicted by the blue, green, and red lines, respectively.
In (a), the line thickness represents the strength of interactions randomly distributed on the different bonds.
In (b), the open circles indicate the vacant sites.
The sites with the orange circles are unpaired sites whose neighbor on the $z$ bond is vacant, which give distinct contributions to the dynamical spin correlation of the $z$ spin component; see Sec.~\ref{sebsec:unpaired}.
(c) Brillouin zone of the honeycomb lattice.
$\bm{K}_a$ and $\bm{K}_b$ denote the reciprocal vectors.
The red dashed lines represent the symmetric lines on which the spin structure factor is plotted in Figs.~\ref{fig_sqw_kmap_random} and \ref{fig_sqw_kmap_imp}.
}
\label{fig_lattice}
\end{center}
\end{figure}

In this study, we focus on the Kitaev model given by~\cite{Kitaev2006}
\begin{align}
  {\cal H}=-\sum_{\gamma=x,y,z}\sum_{\means{jj'}_\gamma}J_{jj'}S_j^\gamma S_{j'}^\gamma,\label{eq:HamilS}
\end{align}
where $S_j^\gamma$ is the $\gamma (=x,y,z)$ component of the $S=1/2$ spin operator at site $j$ on a honeycomb lattice, which consists of two sublattices, $A$ and $B$, as shown in Fig.~\ref{fig_lattice}(a).
The interaction is Ising type and depends on the three directions of the nearest-neighbor (NN) bonds, which is called the Kitaev-type interaction; 
the sums in Eq.~\eqref{eq:HamilS} are taken for the three kinds of the NN bonds, $\means{jj'}_x$, $\means{jj'}_y$, and $\means{jj'}_z$, shown by the blue, green, and red bonds in Fig.~\ref{fig_lattice}(a), respectively.
To introduce disorder, the exchange constant $J_{jj'}$ is assumed to be bond dependent (see Sec.~\ref{subsec:model_disorder}).

The Hamiltonian given in Eq.~\eqref{eq:HamilS} is mapped to a fermionic model by applying the Jordan-Wigner transformation~\cite{PhysRevB.76.193101,PhysRevLett.98.087204,1751-8121-41-7-075001,PhysRevLett.113.197205,PhysRevB.92.115122}.
By introducing the two types of Majorana fermions $c_j$ and $\bar{c}_j$ at each site, the spin operator on the $A$ sublattice is represented by 
\begin{align}
  S_j^x=\frac{1}{2}c_j\tau_j,\quad 
  S_j^y=-\frac{1}{2}\bar{c}_j\tau_j,\quad 
  S_j^z=\frac{i}{2}c_j\bar{c}_j,
\end{align}
and that on the $B$ sublattice is written as
\begin{align}
  S_j^x=\frac{1}{2}\bar{c}_j\tau_j,\quad
  S_j^y=-\frac{1}{2}c_j\tau_j,\quad
  S_j^z=\frac{i}{2}\bar{c}_j c_j,
\end{align}
where $\tau_j=\prod_{j'<j}\left(-2S_{j'}^z\right)$.
Using these relations, the model Hamiltonian is rewritten as
\begin{align}
  {\cal H}=&-\sum_{[jj']_x}\frac{iJ_{jj'}}{4} c_j c_{j'}-\sum_{[jj']_y}\frac{iJ_{jj'}}{4} c_j c_{j'}
  -\sum_{[jj']_z}\frac{J_{jj'}}{4} i c_j c_{j'} \eta_r,
    \label{eq:HamilM}
 \end{align}
where $\sum_{[jj']_\gamma}$ denotes the ordered sum over the NN sites on the $\gamma$ bond with $j\in A$ and $j'\in B$, and $\eta_{r} =i \bar{c}_j \bar{c}_{j'}$ with $r$ being an index for the corresponding $z$ bond $[jj']_z$.
In Eq.~\eqref{eq:HamilM}, $\eta_{r}$ is a $Z_2$ conserved quantity taking $\pm 1$, which can be regarded as a classical variable.  
Hence, Eq.~\eqref{eq:HamilM} is a bilinear Hamiltonian in terms of the $c$ Majorana fermions for a given configuration of $\{\eta_r\}$, and hence, it is easily diagonalized as
\begin{align}
  {\cal H}=\frac{i}{4}\sum_{jj'} c_j A_{jj'}c_{j'}=\sum_{\lambda=1}^{N_{\rm spin}/2} \varepsilon_\lambda \left(f_\lambda^\dagger f_\lambda-\frac{1}{2}\right),\label{eq:hamilM2}
\end{align}
where $A$ is a $N_{\rm spin}\times N_{\rm spin}$ skew symmetric real matrix with $N_{\rm spin}$ being the number of spins (see Sec.~\ref{subsec:model_disorder}), and $\varepsilon_\lambda$ is a positive eigenvalue of $iA$.
Note that the eigenvalues appear in pairs as $\pm\varepsilon_\lambda$; namely, $iA$ is diagonalized by the $N_{\rm spin}\times N_{\rm spin}/2$ matrix $U$ with $U^\dagger U=1$ as
\begin{align}
  \Lambda={\rm diag}\{\varepsilon_1,\varepsilon_2,\cdots,\varepsilon_{N_{\rm spin}/2-1},\varepsilon_{N_{\rm spin}/2}\}
  =U^\dagger iA U.
\label{eq:diagiA}
\end{align}
See also Appendix~\ref{app:zeroT}. 
$f_\lambda$ and $f_\lambda^\dagger$ in Eq.~\eqref{eq:hamilM2} are related with the Majorana fermion operator $c_j$ as 
\begin{align}
  c_j=\sqrt{2}\sum_{\lambda=1}^{N_{\rm spin}/2} \left(U_{j\lambda}f_\lambda +U_{j\lambda}^* f_\lambda^\dagger \right).
\label{eq:c_f}
\end{align}

Thus, Eq.~\eqref{eq:HamilM} indicates that the system is described by the itinerant Majorana fermions $c_j$ and the localized bond variables $\eta_r$. 
For the latter degree of freedom, one can introduce the local variable called flux on each hexagonal plaquette of the honeycomb lattice.
The flux on a plaquette $p$ is defined as
\begin{align}
W_p=2^6 \prod_{j\in p} S_j^{\gamma_j},
\end{align} 
where $\gamma_j(=x,y,z)$ is the type of the bond not belonging to $p$ among the three connected to site $j$.
This is a local conserved quantity taking $\pm 1$, similar to $\eta_r$. 
Indeed, there is a relation between $W_p$ and $\eta_r$ as $W_p=\prod_{r\in p}\eta_r$.
Therefore, the Kitaev spin model in Eq.~\eqref{eq:HamilS} is regarded as a free Majorana fermion problem coupled with the localized fluxes.
This indicates that the quantum spins are fractionalized into the itinerant Majorana fermions and the localized fluxes.

\subsection{Two types of disorder}
\label{subsec:model_disorder}

We consider two types of disorder, bond randomness and site dilution, following the previous work~\cite{Nasu2020disorder}.
In the case of the bond randomness, the exchange constants $J_{jj'}$ are set randomly from a uniform distribution in the range of $[J-\zeta:J+\zeta]$.
The situation is schematically depicted in Fig.~\ref{fig_lattice}(a).
In the case of the site dilution, spins are randomly replaced by vacancies. 
In this case, we set $J_{jj'}=0$ for the bonds including the vacancies and $J_{jj'}=J$ for the remaining bonds, as schematically shown in Fig.~\ref{fig_lattice}(b).
The density of the vacancies is defined as $\rho=N_{\rm vac}/N$, where $N_{\rm vac}$ and $N$ are the numbers of vacancies and sites including the vacancies, respectively; namely, the number of spins remaining in the system is given by $N_{\rm spin}=N-N_{\rm vac}$.

In the case of the site dilution, some spins lack the neighbor on the $z$ bond, as shown by the orange circles in Fig.~\ref{fig_lattice}(b).
We call such sites the unpaired sites, whose contributions to the spin dynamics in the $z$ spin component will be discussed separately in the following sections.

\subsection{Signatures of spin fractionalization}
\label{sec:Signatures-spin-fractionalization}

Before going into the spin dynamics in the presence of disorder, let us briefly review some fundamental properties of the pristine and disordered Kitaev models.
In the pristine Kitaev model without disorder, the signatures of spin fractionalization appear in many physical observables~\cite{PhysRevLett.98.247201,PhysRevLett.112.207203,PhysRevLett.113.187201,Perreault_resonant2016,Halasz2016,Yoshitake2020,Choi_Nonlinear2020}. 
For instance, the specific heat shows two peaks, both of which are crossovers, and half of the entropy $\ln 2$ per spin is released at each crossover~\cite{PhysRevB.92.115122,PhysRevLett.113.197205}.
The lower-$T$ crossover occurring at $T_L\simeq 0.012J$ corresponds to the freezing of all the fluxes $W_p$ to $+1$, and indeed, the temperature scale is set by the excitation gap by flipping $W_p$ from the ground state with all $W_p=+1$. 
Meanwhile, the other crossover occurring at a much higher $T_H\simeq 0.38J$ corresponds to the Fermi degeneracy in the fermions $f$ composed of the itinerant Majorana fermions $c$. 
In terms of the original spin degree of freedom, the crossover at $T_H$ corresponds to the development of the NN spin correlations, which are equivalent to the kinetic energy of the Majorana fermions as shown in the derivation of Eq.~\eqref{eq:HamilM}.
Thus, the temperature scale is in the order of the energy scale of the Kitaev interaction $J$.

The fractionalization manifests itself also in the spin dynamics.
At zero $T$, the dynamical spin correlation exhibits two features.
One is the low-energy sharp peak above a small spin gap caused by the flux excitation and the other is the high-energy continuum ascribed to the excitations of the itinerant Majorana fermions~\cite{PhysRevLett.112.207203,PhysRevB.92.115127}.
At finite $T$, while increasing $T$, the low-energy peak is broadened around $T_L$, and eventually all features including the continuum are smeared above $T_{H}$~\cite{yoshitake2016,Yoshitake2017PRBa,Yoshitake2017PRBb} (see also Sec~\ref{sec:dsf}).
Such behavior has been observed experimentally using the inelastic neutron scattering measurements as the evidence of the spin fractionalization in the Kitaev magnets~\cite{banerjee2016proximate,Do2017majorana,Banerjee2017,banerjee2018excitations}.

\subsection{Disorder effects on thermodynamics and transport}
\label{sec:thermodynamics-transport}

The thermodynamic properties in the presence of the two types of disorder were recently studied by the authors in Ref.~\cite{Nasu2020disorder}.
The higher-$T$ crossover at $T_H$ does not show significant changes for both types of disorder, while the specific heat peak is slightly shifted to higher (lower) $T$ for the bond randomness (site dilution).
On the other hand, the lower-$T$ crossover at $T_L$ shows contrasting behavior for the two types of disorder.
In the case of the bond randomness, $T_L$ is shifted to the lower-$T$ side with increasing the strength of disorder, while the entropy from the fluxes are almost fully released at the lowest $T$.
In the case of the site dilution, however, $T_L$ is nearly unchanged but the peak of the specific heat is suppressed by increasing the density of vacancies.
This implies that the $\frac12 \ln 2$ entropy of the fluxes are not fully released in the crossover at $T_L$ and $W_p$ remains fluctuating even below this $T$.
This is presumably due to the smaller excitation gap for the fluxes defined for larger plaquettes including the vacancy sites~\cite{Willans2010}, and the entropy is expected to be fully released at the extremely lower $T$.

Contrasting responses to the two types of disorder were also observed in the thermal transport properties~\cite{Nasu2020disorder}.
While the longitudinal thermal conductivity is suppressed in a similar manner by the two types of disorder, the thermal Hall conductivity $\kappa^{xy}$ induced by a magnetic field exhibits contrasting $T$ dependences at low $T$:
The half-quantized plateau of $\kappa^{xy}/T$, which is a hallmark of the topological gapped state in the magnetic field~\cite{Kitaev2006}, remains robust against the weak bond randomness but it is fragile against the introduction of the site dilution, closely correlated with the disorder effects on the flux excitations discussed above.
The results may be relevant to the recent experiments suggesting that the quality of samples plays a key role for the quantization of $\kappa^{xy}/T$~\cite{kasahara2018majorana,Yokoi2020pre}.

\section{Method}
\label{sec:method}

\subsection{Monte Carlo simulation}\label{sec:MC-simulations}

To calculate the $T$ dependence of the physical quantities, we apply the quantum MC simulation to the bilinear Hamiltonian in terms of the Majorana fermions in Eq.~\eqref{eq:hamilM2}~\cite{PhysRevLett.113.197205,PhysRevB.92.115122}.
The partition function of the system for a given configuration of $\{\eta_r\}$ is written as
\begin{align}
  Z_{\{\eta_r\}}={\rm Tr}_{\{c_j\}}e^{-\beta {\cal H}}=\prod_{\lambda=1}^{N_{\rm spin}/2} 2\cosh \frac{\beta \varepsilon_{\lambda}}{2},
\end{align}
and the free energy is given by
\begin{align}
  F_{\{\eta_r\}} = -\frac{1}{\beta}\ln Z_{\{\eta_r\}},
\end{align}
where $\beta=1/T$ is the inverse temperature (we set the Boltzmann constant $k_{\rm B}=1$).
Note that the partition function of the whole system is written as $Z=\sum_{\{\eta_r\}}Z_{\{\eta_r\}}$.
In the MC simulations, a sequence of configurations of $\{\eta_r\}$, $(\{\eta_r\}_1, \{\eta_r\}_2, \cdots, \{\eta_r\}_{N_{\rm MC}})$, is generated so as to reproduce the distribution of $e^{-\beta F_{\{\eta_r\}}}$.
Then, the thermal average of an operator ${\cal O}$ is evaluated as the MC average:
\begin{align}
  \means{{\cal O}}
  =\frac{1}{Z}\sum_{\{\eta_r\}}e^{-\beta F_{\{\eta_r\}}}\means{{\cal O}}_{\{\eta_r\}} \simeq 
  \frac{1}{N_{\rm MC}} \sum_{l=1}^{N_{\rm MC}} \means{{\cal O}}_{\{\eta_r\}_l},
\end{align}
where $\means{{\cal O}}_{\{\eta_r\}}$ is the expectation value of ${\cal O}$ for the configuration $\{\eta_r\}$.
Hereafter, we write $\means{{\cal O}}_{\{\eta_r\}}$ as $\means{{\cal O}}_\eta$ for simplicity;
it is calculated from
\begin{align}
 \means{{\cal O}}_\eta = \frac{1}{Z_{\{\eta_r\}}}{\rm Tr}_{\{c_j\}}{\cal O} e^{-\beta {\cal H}}.
\end{align}

The numerical calculations are performed on the cluster including $N=2L^2$ sites with $L=12$, where the shifted periodic boundary condition is imposed (see Ref.~\cite{Nasu2020disorder} for the details).
We prepare 20 configurations of $\{J_{jj'}\}$ for the bond randomness and 10 configurations of vacancies for the site dilution.
In each configuration of disorder, we generate a sequence of configurations of $\{\eta_r\}$ using the Markov chain MC simulation and obtain 20~000 samples of $\{\eta_r\}$ after 10~000 MC steps for thermalization.
Among the samples, we pick up 100 samples every 200 for measurement of physical quantities.
The errors are evaluated from the standard deviations calculated for the MC averages in different configurations of the disorders.
Meanwhile, in the pristine case, they are evaluated for the MC averages in 20 independent runs.

In the case of the site dilution, the vacancies are distributed randomly, and hence, the number of the vacancies on the $A$ sublattice, $N_A$, is not always the same as that on the $B$ sublattice, $N_B$, in each random sample.
Since the vacancies induce the zero energy excitations called the Majorana zero modes whose number is proportional to $|N_A-N_B|$~\cite{Willans2011,Santhosh2012}, the case with $N_A=N_B$ is rather special.
The case with $N_A=N_B$ ($N_A\neq N_B$) is called the compensated (uncompensated) case.

\subsection{Dynamical spin correlation}
\label{sec:dynamical-spin-correlation}

In this section, we describe how to evaluate the dynamical spin correlation functions by using the MC sampling.
The Majorana fermion representation introduced in Sec.~\ref{subsec:model_Majorana} allows us to evaluate the $S^z$ component of the dynamical spin correlations.
Although they were computed by quantum MC techniques based on the path-integral framework in the previous studies~\cite{yoshitake2016,Yoshitake2017PRBa,Yoshitake2017PRBb}, here we use an alternative method based on the matrix representation in the real-time domain~\cite{Udagawa2018}.
This method has the advantage of not requiring an analytic continuation.
In the real-time framework, the onsite and NN $z$-bond components are represented as
\begin{align}
  \means{S_j^z(t) S_j^z}_{\eta}&=\frac{1}{4}\means{e^{i{\cal H}t} c_j e^{-i{\cal H}^{(r)}t} c_{j} }_{\eta}\nonumber\\
&  =\frac{1}{2}\sqrt{{\rm det} C(t)}\left[C^{-1}(t) C'(t)\right]_{jj},\label{eq:corr_orig1}\\
  \means{S_j^z(t) S_{j'}^z}_{\eta}&=\frac{i\eta_r}{4}\means{e^{i{\cal H}t} c_j e^{-i{\cal H}^{(r)}t}  c_{j'}}_{\eta}\nonumber\\
 & =\frac{i\eta_r}{2}\sqrt{{\rm det} C(t)}\left[C^{-1}(t) C'(t)\right]_{j'j}\label{eq:corr_orig2},
\end{align}
respectively, where the $N_{\rm spin}\times N_{\rm spin}$ matrices $C(t)$ and $C'(t)$ are given by
\begin{align}
  C(t)&=(1+e^{-\beta iA})^{-1} (1+e^{-(\beta-it) iA}e^{-it iA^{(r)}}),\\
  C'(t)&=(1+e^{-\beta iA})^{-1}e^{-(\beta-it) iA},
\end{align}
respectively.
Here, ${\cal H}^{(r)}=\frac{i}{4}\sum_{j,j'}c_j A_{jj'}^{(r)} c_{j'}$ is the Hamiltonian where $\eta_r$ on the bond is flipped, and $A^{(r)}$ is the corresponding skew matrix; in Eq.~\eqref{eq:corr_orig2}, $j$ and $j'$ are the $A$ and $B$ sublattice sites on a $z$ bond $r$, respectively. 
Note that the dynamical spin correlations for further neighbors beyond the NN sites as well as the NN correlations of the $S^x$ and $S^y$ components on the $z$ bonds are zero even in the presence of disorder because of the existence of the local conserved quantities~\cite{PhysRevLett.98.247201}.
By introducing
\begin{align}
X=(1+e^{-\beta iA})^{-1}, 
\quad Y(t)=e^{-it iA^{(r)}},\label{eq:XYforms}
\end{align}
$C(t)$ is related with $C'(t)$ as 
\begin{align}
  C(t)=X+C'(t) Y(t).
\end{align}
The matrix element of $X$ and $Y(t)$ are calculated as 
\begin{align}
  X_{jj'}&=\sum_{\lambda=1}^{N_{\rm spin}/2} 
  \left(U_{j\lambda}U_{j'\lambda}^* [1-f(\varepsilon_{\lambda})] 
  +U_{j\lambda}^* U_{ j'\lambda} f(\varepsilon_{\lambda})\right),\label{eq:Xform}\\
  Y_{jj'}(t)&=\sum_{\lambda=1}^{N_{\rm spin}/2} 
  \left(U_{j\lambda}^{(r)}U_{j'\lambda}^{(r)*} e^{-it \varepsilon_{\lambda}^{(r)}}
  +U_{j\lambda}^{(r)*} U_{ j'\lambda}^{(r)}e^{it \varepsilon_{\lambda}^{(r)}}  \right),\label{eq:Yform}
\end{align}
where $f(\varepsilon)=1/(e^{\beta \varepsilon}+1)$ is the Fermi distribution function; $\varepsilon_\lambda^{(r)}$ is the positive eigenvalue of $iA^{(r)}$, which is diagonalized by the $N_{\rm spin}\times N_{\rm spin}/2$ matrix $U^{(r)}$ [see Eq.~\eqref{eq:diagiA}].
In a similar manner, the matrix element of $C'(t)$ is obtained as
\begin{align}
  C'_{jj'}(t)&=\sum_{\lambda=1}^{N_{\rm spin}/2} 
  \left(U_{j\lambda}U_{j'\lambda}^* f(\varepsilon_{\lambda}) e^{it \varepsilon_{\lambda}}
  +U_{j\lambda}^* U_{ j'\lambda}e^{-it \varepsilon_{\lambda}}[1-f(\varepsilon_{\lambda})]  \right).\label{eq:Cp}
\end{align}
We note that Eqs.~\eqref{eq:Xform}, \eqref{eq:Yform}, and \eqref{eq:Cp} allow the stable calculations of the dynamical spin correlations down to low $T$;
the functional forms in the limit of $T\to 0$ are discussed in Appendix~\ref{app:zeroT}.

As $C(t=0)=1$, the equal-time spin correlations are obtained as 
\begin{align}
\label{eq:SjSj_t=0}
  \means{S_j^z S_j^z}_{\eta}&=\frac{1}{4},\\
\label{eq:SjSj'_t=0}
  \means{S_j^z S_{j'}^z}_{\eta}&=\frac{i\eta_r}{2}\sum_{\lambda=1}^{N_{\rm spin}/2} 
  \left( U_{j'\lambda}U_{j\lambda}^* f(\varepsilon_{\lambda}) + U_{j'\lambda}^* U_{ j\lambda} [1-f(\varepsilon_{\lambda})] \right).
\end{align}
In the time evolution for $t>0$, one needs to fix the phase of $\sqrt{{\rm det}C(t)}$ in Eqs.~\eqref{eq:corr_orig1} and \eqref{eq:corr_orig2}. 
In the practical calculations, starting from the equal-time spin correlations in Eqs.~\eqref{eq:SjSj_t=0} and \eqref{eq:SjSj'_t=0}, we determine the phase sequentially so as to make the correlation functions continuous.

\subsection{Dynamical spin correlation for unpaired spins}
\label{sebsec:unpaired}

The dynamical spin correlations in Eqs.~\eqref{eq:corr_orig1} and \eqref{eq:corr_orig2} are defined when two spins are present on the $z$ bond $r$.
In the case of the site dilution, this does not hold for all the $z$ bonds; 
there are some spins whose counterpart on the $z$ bond is replaced by a vacancy. 
We call such spins the unpaired spins. 
For the unpaired spins, the NN correlation is zero and the onsite one is explicitly given as 
\begin{align}
  \means{S_j^z(t) S_j^z}_\eta=\frac{1}{2}\sum_{\lambda=1}^{N_{\rm spin}/2} 
  |U_{j\lambda}|^2  \left(
    e^{it \varepsilon_{\lambda}} f(\varepsilon_{\lambda})
  + e^{-it \varepsilon_{\lambda}}[1-f(\varepsilon_{\lambda})]\right),
  \label{eq:Sunp}
\end{align}
because the exchange interaction on the $z$ bond is absent and thereby ${\cal H}^{(r)}$ in Eq.~\eqref{eq:corr_orig1} is identical to ${\cal H}$~\cite{Udagawa2018}.

\subsection{Dynamical spin structure factor}
\label{sec:dynamical-spin-structure-factor}

The dynamical spin correlation function as a function of the frequency is defined by the Fourier transformation with respect to time $t$ as
\begin{align}
  {\cal S}_{jj'}^{zz}(\omega) = \frac{1}{2\pi}\int_{-\infty}^{\infty}\means{S_j^z(t) S_{j'}^z}e^{i\omega t} dt.
\end{align}
As we evaluate the correlation functions for $t\geq 0$, we use the following relation instead of the above equation:
\begin{align}
  {\cal S}_{jj'}^{zz}(\omega) = \frac{1}{\pi}{\rm Re}\int_{0}^{\infty}\means{S_j^z(t) S_{j'}^z}e^{i\omega t} dt.
\end{align}

As mentioned before, the spin correlations are nonzero only for the onsite and NN sites on the $z$ bond. 
We compute these two components separately as
\begin{align}
\label{eq:SLO}
  {\cal S}^{{\rm onsite}}(\omega) = \frac{1}{N}\sum_j{\cal S}_{jj}^{zz}(\omega), \\ 
\label{eq:SNN}
  {\cal S}^{{\rm NN}}(\omega) = \frac{2}{N}\sum_{\means{jj'}_z}{\cal S}_{jj'}^{zz}(\omega),
\end{align}
for the case of the bond randomness.
In the case of the site dilution, we classify the lattice sites into three types: the vacancy sites $j\in{\cal G}_{\rm vac}$, the unpaired sites $j\in{\cal G}_{\rm unpair}$, and the others $j\in{\cal G}_{\rm pair}$, whose numbers of sites are given by $N_{\rm vac}$, $N_{\rm unpair}$, and $N_{\rm pair}$, respectively; $N=N_{\rm vac} + N_{\rm unpair} + N_{\rm pair}$ and $N_{\rm spin} = N_{\rm unpair} + N_{\rm pair}$. 
From Eq.~\eqref{eq:Sunp}, we define the onsite dynamical spin correlation for the unpaired spins as
\begin{align}
  &
{\cal S}_{\rm unpair}(\omega) = \frac{1}{N_{\rm unpair}}\sum_{j\in {\cal G}_{\rm unpair}}{\cal S}_{jj}^{zz}(\omega)\nonumber\\
  &= \frac{1}{2N_{\rm unpair}}\sum_{j\in {\cal G}_{\rm unpair}}\sum_{\lambda=1}^{N_{\rm spin}/2}
  |U_{j\lambda}|^2  \left[
    f(\varepsilon_{\lambda})\delta(\omega+\varepsilon_{\lambda})\right.\nonumber\\
& \qquad\qquad\qquad\qquad\qquad\qquad \left.   +
    f(-\varepsilon_{\lambda})\delta(\omega-\varepsilon_{\lambda})\right].\label{eq:Sunpw}
\end{align}
Here, we omit the superscript ``onsite'' as the NN component is automatically zero for the unpaired spins. 
Equation~\eqref{eq:Sunpw} indicates that ${\cal S}_{\rm unpair}(\omega)$ at $T=0$ coincides with the local density of states of the Majorana fermions at the unpaired sites.
Then, for the case of the site dilution, the onsite and NN dynamical spin correlations are given by
\begin{align}
  {\cal S}^{\rm onsite}(\omega) &= \frac{N_{\rm unpair}}{N}{\cal S}_{\rm unpair}(\omega) + \frac{N_{\rm pair}}{N}{\cal S}_{\rm pair}^{\rm onsite}(\omega)
,\label{eq:SLOw}\\
  {\cal S}^{\rm NN}(\omega) &= \frac{N_{\rm pair}}{N}{\cal S}_{\rm pair}^{\rm NN}(\omega),
\label{eq:SNNw}
\end{align}
where ${\cal S}_{\rm pair}^{\rm onsite}(\omega)$ and ${\cal S}_{\rm pair}^{\rm NN}(\omega)$ are the onsite and NN correlations for the bonds with no vacancies defined in a similar manner to Eqs.~\eqref{eq:SLO} and \eqref{eq:SNN} as
\begin{align}
  {\cal S}_{\rm pair}^{\rm onsite}(\omega) &= \frac{1}{N_{\rm pair}}\sum_{j\in {\cal G}_{\rm pair}}{\cal S}_{jj}^{zz}(\omega),\label{eq:Slocw}\\
  {\cal S}_{\rm pair}^{\rm NN}(\omega) &= \frac{2}{N_{\rm pair}}\sum_{\means{jj'}_z \in {\cal G}_{\rm pair}}{\cal S}_{jj'}^{zz}(\omega),\label{eq:Snnw}
\end{align}
respectively.

The dynamical spin structure factor is obtained by the Fourier transformation with respect to the real-space position as 
\begin{align}
  {\cal S}(\bm{q},\omega)=\frac{1}{N}\sum_{jj'}{\cal S}_{jj'}(\omega) e^{-i\bm{q}\cdot(\bm{r}_j-\bm{r}_{j'})},
\end{align}
where ${\cal S}_{jj'}(\omega)$ is defined as
\begin{align}
  {\cal S}_{jj'}(\omega)=\frac{1}{3}\left[
    {\cal S}_{jj'}^{xx}(\omega)+{\cal S}_{jj'}^{yy}(\omega)+{\cal S}_{jj'}^{zz}(\omega)\right]. 
\end{align}
In the present calculation, we assume ${\cal S}_{jj'}(\omega)={\cal S}_{jj'}^{zz}(\omega)$ for simplicity.
Alternatively, by using the fact that the spin correlations are limited to onsite and NN, ${\cal S}(\bm{q},\omega)$ can be written as
\begin{align}
  {\cal S}(\bm{q},\omega)={\cal S}^{\rm onsite}(\omega) + c_{\bm{q}}{\cal S}^{\rm NN}(\omega)\label{eq:Sqw}
\end{align}
where
\begin{align}
c_{\bm{q}}=\frac{1}{3}\left(2\cos\frac{q_x}{2}\cos\frac{q_y}{2\sqrt{3}}+\cos\frac{q_y}{\sqrt{3}}\right).
\end{align}
In Eq.~\eqref{eq:Sqw}, we use Eqs.~\eqref{eq:SLO} and \eqref{eq:SNN} for the bond randomness and Eqs.~\eqref{eq:SLOw} and \eqref{eq:SNNw} for the site dilution.
To reduce the calculation cost, the site summations in Eqs.~\eqref{eq:SLO} and \eqref{eq:SNN} [Eqs.~\eqref{eq:Slocw} and \eqref{eq:Snnw}] are approximately calculated by the random averages over the twelve $z$ bonds chosen randomly from $N/2$ ($N_{\rm pair}/2$) bonds for the former (latter).
On the other hand, the summation in ${\cal S}_{\rm unpair}(\omega)$ given in Eq.~\eqref{eq:Sunpw} is taken for all unpaired sites.

We also discuss the static spin correlations for comparison. 
They are measured by the spin structure factor defined as
\begin{align}
S(\bm{q}) = \frac1N \sum_{jj'} \langle S_j^z S_{j'}^z \rangle
e^{-i\bm{q}\cdot(\bm{r}_j-\bm{r}_{j'})}. 
\end{align}
This quantity is given by the $\omega$ integral of ${\cal S}(\bm{q},\omega)$ as 
\begin{align}
  S(\bm{q})=\int_{-\infty}^\infty {\cal S}(\bm{q},\omega) d\omega.
\label{eq:Sq}
\end{align}

\subsection{Magnetic susceptibility}
\label{sec:Magnetic-susceptibility}

We also compute the isothermal magnetic susceptibility defined as
\begin{align}
  \chi=\frac{1}{N}\sum_{jj'}\int_0^\beta d\tau \means{e^{\tau {\cal H}} S_j^z e^{-\tau {\cal H}} S_{j'}^z}
\label{eq:susceptibility}
\end{align}
This is related with the dynamical structure factor by the fluctuation-dissipation theorem as
\begin{align}
  \chi=\int_{-\infty}^\infty d\omega' 
  \frac{(1-e^{-\beta \omega'}){\cal S}(\bm{q}=0,\omega')}{\omega'}.
\end{align}
In the following analysis, we decompose it into two parts as 
$\chi = \chi^{\rm reg}+ \chi^{\rm Curie}$:
\begin{align}
  \chi^{\rm reg}&={\cal P}\int_{-\infty}^\infty d\omega' 
  \frac{(1-e^{-\beta \omega'}){\cal S}(\bm{q}=0,\omega')}{\omega'}\label{eq:suscepadiabatic}\\
  \chi^{\rm Curie}&=\frac{1}{T}\lim_{\epsilon\to 0}\int_{-\epsilon}^\epsilon d\omega' 
  {\cal S}(\bm{q}=0,\omega'),\label{eq:suscepcurie}
\end{align}
where ${\cal P}$ denotes the principal integral, ${\cal P}\int_{-\infty}^\infty=\lim_{\epsilon\to 0}\left(\int_{-\infty}^{-\epsilon}+\int_{\epsilon}^\infty\right)$. 
In the numerical calculations, we set $\epsilon/J=7.5\times 10^{-4}$.
Note that $\chi^{\rm reg}$ corresponds to the adiabatic magnetic susceptibility, and $\chi^{\rm Curie}$ describes the Curie contribution.

\subsection{NMR relaxation rate}
\label{sec:NMR-relaxation-rate}

In addition, we calculate the NMR relaxation rate, which is given by
\begin{align}
\frac{1}{T_1}\propto \sum_{\bm{q}}\left|A_{\bm{q}}\right|^2 {\cal S}(\bm{q},\omega_0),
\end{align}
where $A_{\bm{q}}$ is the hyperfine coupling constant and $\omega_0$ is the resonant frequency in the NMR measurement, which is much smaller compared to the other energy scales.
Since $A_{\bm{q}}$ is dependent in each material, following Ref.~\cite{yoshitake2016,Yoshitake2017PRBa,Yoshitake2017PRBb}, we calculate the onsite and NN contributions separately:
\begin{align}
\label{eq:NMRrate_onsite}
\frac{1}{T_1^{\rm onsite}}&={\cal S}^{\rm onsite}(\omega)\Big|_{\omega \to 0},\\
\frac{1}{T_1^{\rm NN}}&={\cal S}^{\rm NN}(\omega)\Big|_{\omega \to 0}.
\label{eq:NMRrate_NN}
\end{align}
Note that ${\cal S}_{jj'}(\omega)$ at $\omega=0$ does not contribute to the NMR relaxation rate unlike the isothermal magnetic susceptibility.

In the present calculations, we do not introduce an external magnetic field, while the NMR experiments are usually done at finite magnetic fields. 
Thus, our results should be compared with the zero-field data, which are obtained by the nuclear quadrupole resonance~\cite{Nagai2020}.

\section{Result}
\label{sec:result}

\subsection{Dynamical spin structure factor}
\label{sec:dsf}

\begin{figure}[t]
  \begin{center}
  \includegraphics[width=\columnwidth,clip]{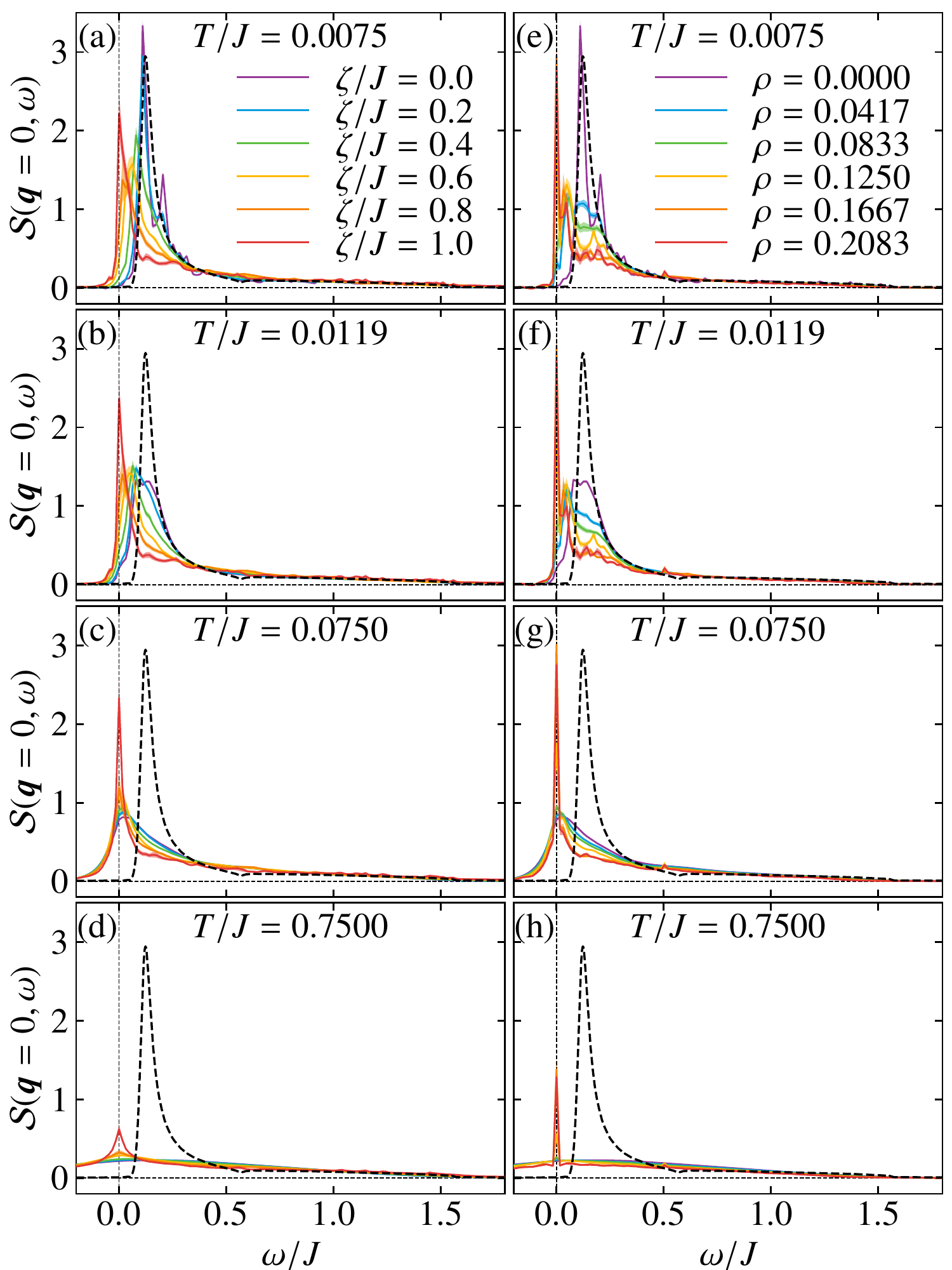}
  \caption{
Frequency dependences of the dynamical spin structure factor at $\bm{q}=0$ for the disordered Kitaev model while changing (a)--(d) the bond randomness $\zeta/J$ and (e)--(h) the site dilution $\rho$: (a)(e) $T/J=0.0075$, (b)(f) $T/J=0.0119$, (c)(g) $T/J=0.0750$, and (d)(h) $T/J=0.7500$.
The errors of the calculations are shown as the shades for each data.
The dashed lines represent the the result for the pristine Kitaev model at $T=0$ in the thermodynamic limit.
  }
\label{fig_sqw_zero}
\end{center}
\end{figure}
  
\begin{figure*}[t]
  \begin{center}
  \includegraphics[width=2\columnwidth,clip]{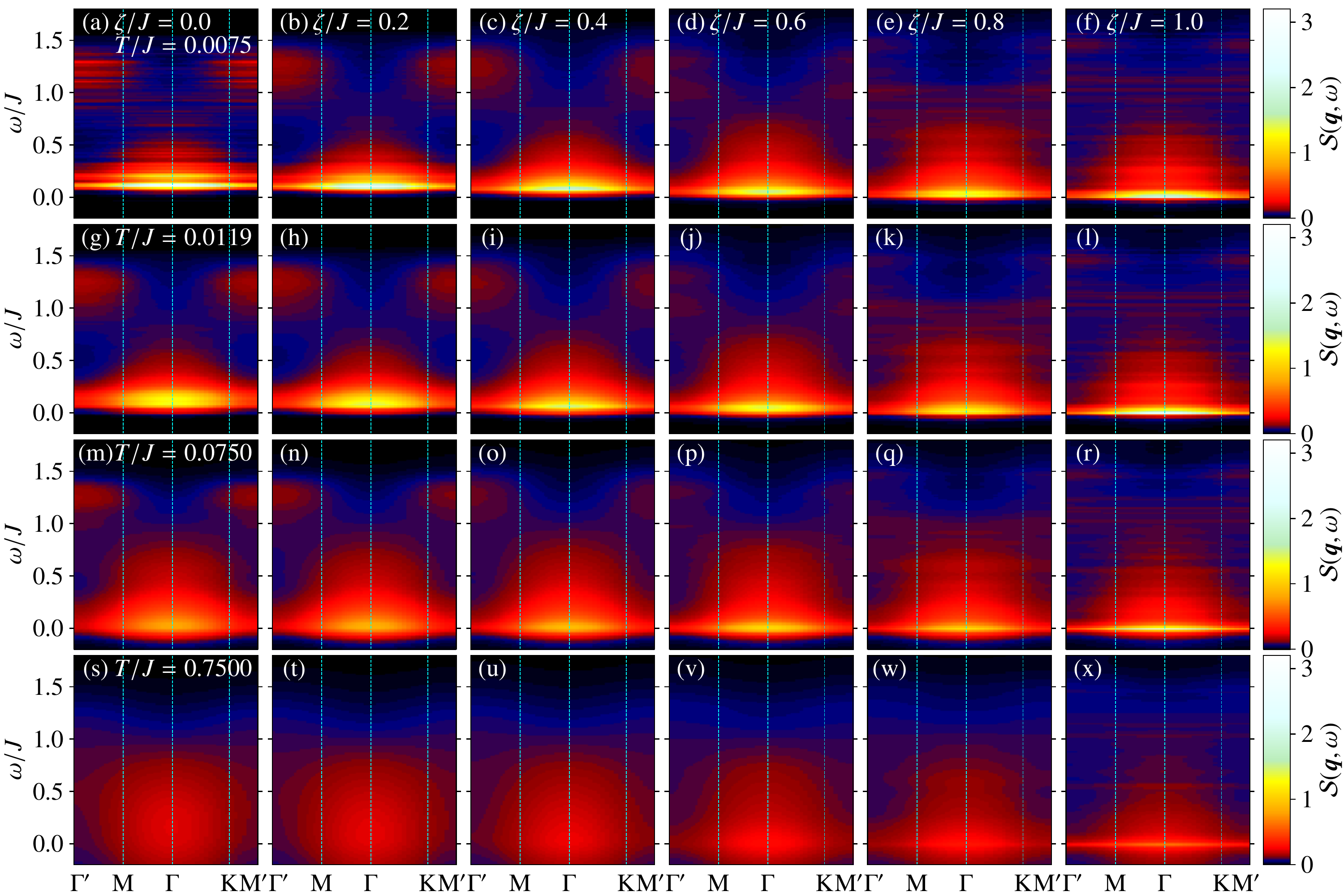}
  \caption{
  Dynamical spin structure factor for the system with the bond randomness.
  (a)--(f), (g)--(l), (m)--(r), and (s)--(x) correspond to the results at $T/J=0.0075$, $0.0119$, $0.075$, and $0.75$, respectively, while changing $\zeta/J$.
  }
\label{fig_sqw_kmap_random}
\end{center}
\end{figure*}

\begin{figure*}[t]
  \begin{center}
  \includegraphics[width=2\columnwidth,clip]{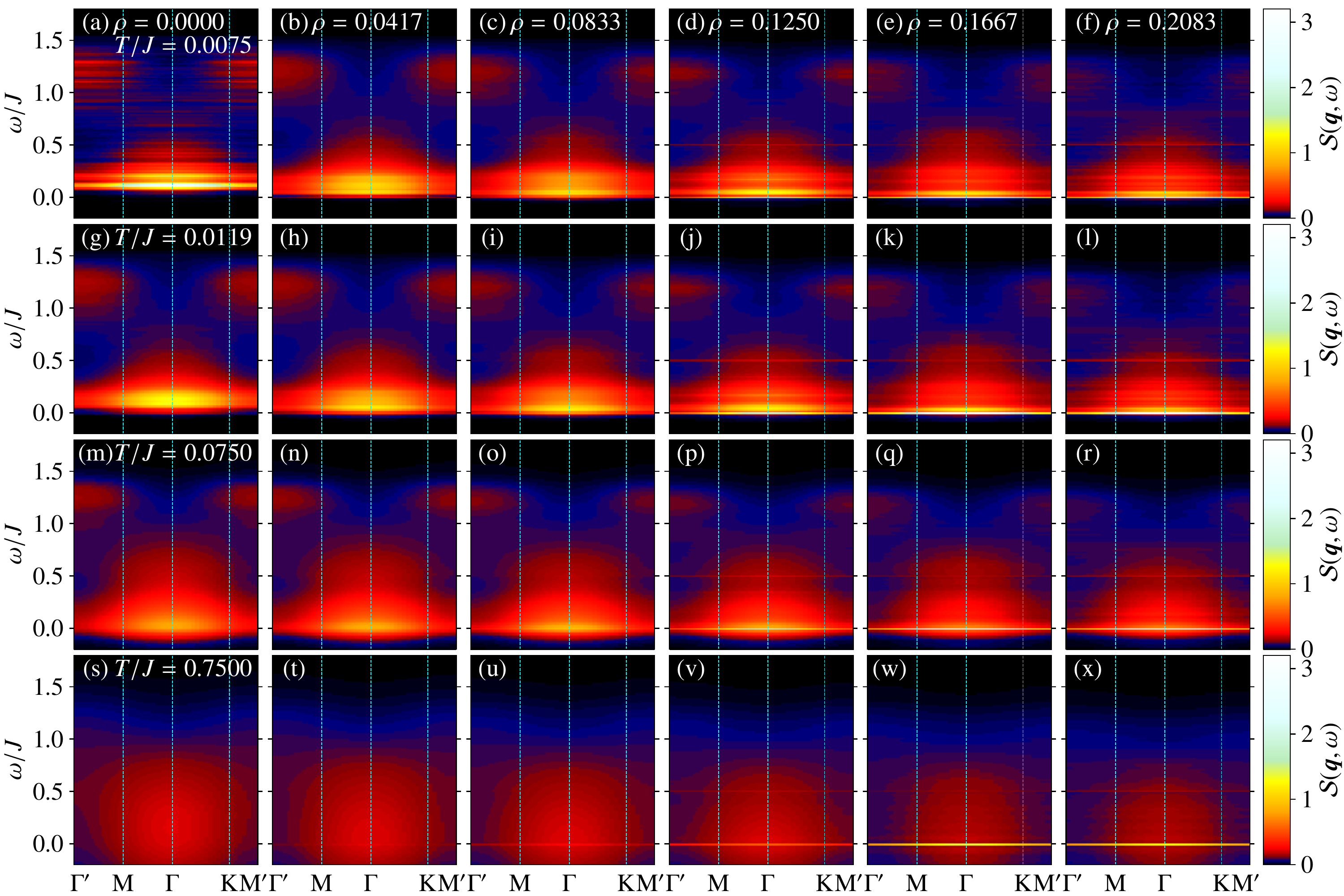}
  \caption{
Dynamical spin structure factor for the system with the site dilution.
(a)--(f), (g)--(l), (m)--(r), and (s)--(x) correspond to the results at $T/J=0.0075$, $0.0119$, $0.075$, and $0.75$, respectively, while changing $\rho$.
}
\label{fig_sqw_kmap_imp}
\end{center}
\end{figure*}

First, we present the quantum MC results for the dynamical spin structure factor calculated by Eq.~\eqref{eq:Sqw}.
Figure~\ref{fig_sqw_zero} shows the frequency dependences of ${\cal S}(\bm{q},\omega)$ at $\bm{q}=0$ while changing the disorder strength at several $T$.
The detailed data decomposed into the onsite and NN contributions are given in Appendix~\ref{app:detail-Sw}.

In the pristine case without disorder, there is a spin gap corresponding to the lowest-energy excitation by flipping the neighboring two fluxes, and accordingly, the spectra shows a single sharp peak at $\omega/J\sim 0.2$ above the gap.
In the higher-energy region beyond the low-energy peak, a broad structure appears up to $\omega/J\sim 1.5$, which dominantly comes from the excitation of itinerant Majorana fermions whose bandwidth is $1.5J$.
Such behaviors in the ground state in the thermodynamic limit of $N\to \infty$ are shown by the dashed lines in Fig.~\ref{fig_sqw_zero}~\cite{PhysRevLett.112.207203,PhysRevB.92.115127} (see also Appendix~\ref{app:zeroT}).
The purple lines in Fig.~\ref{fig_sqw_zero} show our numerical data for the $N=2L^2$ cluster with $L=12$.
At the lowest $T/J=0.0075$, the result well reproduces the two features as shown in Figs.~\ref{fig_sqw_zero}(a) and \ref{fig_sqw_zero}(e), while the peak at $\omega/J\sim 0.2$ is split into two, presumably due to the finite-size effect; see below.

When we introduce the bond randomness, the low-energy peak is shifted to the lower-energy side, and eventually, the peak reaches $\omega=0$ at $\zeta/J\sim1.0$, as shown in Fig.~\ref{fig_sqw_zero}(a).
The result suggests the reduction and closing of the spin gap.
This behavior appears to well correlate with the collapse of the specific heat peak and the half-quantized plateau in the thermal Hall conductivity found in the previous study~\cite{Nasu2020disorder} (see also Sec.~\ref{sec:thermodynamics-transport}).
On the other hand, in the case of the site dilution, the peak is suppressed without noticeable energy shift, but instead, several lower-energy peaks inside the spin gap, including the zero-energy one, are developed while increasing the density of vacancies, $\rho$, as shown in Fig.~\ref{fig_sqw_zero}(e).
This appears to also be related with the previous results for the thermodynamics and transport; in particular, the collapse of the half-quantized plateau of the thermal Hall conductivity for the site dilution~\cite{Nasu2020disorder}.
Thus, the bond randomness and the site dilution cause qualitatively different behaviors in the low-energy spectra of the dynamical spin structure factor:
The former leads to the decrease and closing of the spin gap, while the latter immediately collapses the spin gap by the formation of zero-energy states.

Let us comment on the zero-energy peak appearing for the site dilution. 
This is associated with the zero-energy modes, whose number is proportional to the uncompensated vacancies, $|N_A-N_B|$, as mentioned in Sec~\ref{sec:MC-simulations}. 
Hence, it comes from the uncompensated samples with $N_A\neq N_B$. 
The probability distribution of the realization of $(N_A,N_B)$ as a function of $N_A$ for a fixed $N_A+N_B$ takes a sharp peak at $N_A=N_B$, which corresponds to the compensated case; the peak height diverges and the peak width goes to zero in the thermodynamic limit. 
This naively suggests that the contributions from the uncompensated cases will vanish in the thermodynamic limit, and hence, the zero-energy peak also vanishes. 
However, the absence or persistence of the zero-energy peak remains as an unsettled issue, as discussed for electrons in diluted graphene~\cite{Pereira2006,Pereira2008,ShangduanWu2008,Hafner2014}. 
We show the analysis of our data on this issue in Appendix~\ref{app:compensate}.

The $T$ evolution of ${\cal S}(\bm{q}=0,\omega)$ for the case of the bond randomness is shown in Figs.~\ref{fig_sqw_zero}(a)--\ref{fig_sqw_zero}(d).
The low-energy peak is strongly suppressed around $T=T_L\simeq 0.012J$ where the crossover related to the fluxes takes place~\cite{Nasu2020disorder}, but the high-energy broad structure remains largely intact.
On the other hand, in the case of the site dilution, while the overall behavior looks similar as shown in Figs.~\ref{fig_sqw_zero}(e)--\ref{fig_sqw_zero}(h), the zero-energy peak remains more robustly even for $T> T_H$ compared to the case of the strong bond randomness.
In addition, we note that a small peak appears at $\omega/J=0.5$ for large $\rho$; see  Appendix~\ref{app:detail-Sw}.

Next, we examine the dynamical structure factor including the $\bm{q}$ dependence.
Figure~\ref{fig_sqw_kmap_random} shows ${\cal S}(\bm{q},\omega)$ in the case of the bond randomness at several $\zeta$ and $T$. 
The $\bm{q}$ dependences are plotted along the symmetric lines shown by the red dashed lines in Fig.~\ref{fig_lattice}(c). 
In the pristine case with $\zeta/J=0$, as shown in Fig.~\ref{fig_sqw_kmap_random}(a), the low-energy peak shows a weak $\bm{q}$ dependence with a maximum at the $\Gamma$ point, while the high-energy broad structure has a rather weak intensity around this point at the lowest $T$, forming an hour-glass-like continuum.
The horizontal stripes are an artifact due to the finite-size effect as it is absent in the simulations in larger clusters~\cite{Yoshitake2017PRBb} [the peak splitting in Figs.~\ref{fig_sqw_zero}(a) and \ref{fig_sqw_zero}(e) has the same origin].
The $T$ evolution of ${\cal S}(\bm{q},\omega)$ in the pristine Kitaev model shown in Figs.~\ref{fig_sqw_kmap_random}(a), \ref{fig_sqw_kmap_random}(g), \ref{fig_sqw_kmap_random}(m), and \ref{fig_sqw_kmap_random}(s) is consistent with the previous studies~\cite{yoshitake2016,Yoshitake2017PRBa,Yoshitake2017PRBb,Udagawa2018,Yoshitake2020}:
The low-energy peak is smeared out above $T_L$ and the hour-glass-like magnetic continuum is left in the intermediate-$T$ region for $T_L \lesssim T \lesssim T_H$ [Fig.~\ref{fig_sqw_kmap_random}(m)], which is eventually smeared out by a further increase of $T$ [Fig.~\ref{fig_sqw_kmap_random}(s)]. 
These behaviors are consistent with the results observed in the inelastic neutron scattering measurement for a candidate materials $\alpha$-RuCl$_3$~\cite{Do2017majorana}.

By introducing the bond randomness, the dynamical spin structure factor changes significantly.
The $\zeta$ dependence at $T/J=0.0075$ is shown in Figs.~\ref{fig_sqw_kmap_random}(a)--\ref{fig_sqw_kmap_random}(f).
In the low-energy region, the spin gap is closed by introducing disorder as seen in Fig.~\ref{fig_sqw_zero}(a); the low-energy strong peak, which originates from the softening of the coherent peak, is observed at $\omega \simeq0$, as shown in Figs.~\ref{fig_sqw_kmap_random}(e) and \ref{fig_sqw_kmap_random}(f).
On the other hand, the $\bm{q}$ dependence at high energy appears to retain the hour-glass-like continuum in the pristine case.
The result indicates that the bond randomness affects the flux excitations dominantly rather than the Majorana fermion excitations. 

While increasing $T$, the hour-glass-like continuum remains to be observed for all $\zeta/J$ below $T\simeq T_H$, although it is overall weakened with increasing $\zeta/J$ [Figs.~\ref{fig_sqw_kmap_random}(g)-\ref{fig_sqw_kmap_random}(r)]. 
Carefully looking the data, however, we note that the intensities at $\bm{q}$ away from the $\Gamma$ point are suppressed in the high-energy region but instead increased around $\omega=0$ while increasing $\zeta$.
As the result, a strong peak with weak $\bm{q}$ dependence is left at $\omega\simeq0$, as shown in Figs.~\ref{fig_sqw_kmap_random}(r) and \ref{fig_sqw_kmap_random}(x); see also Figs.~\ref{fig_sqw_zero}(c) and \ref{fig_sqw_zero}(d). 

We show ${\cal S}(\bm{q},\omega)$ for the case of the site dilution in Fig.~\ref{fig_sqw_kmap_imp}.
The results for the pristine case with $\rho=0$ in Figs.~\ref{fig_sqw_kmap_imp}(a), \ref{fig_sqw_kmap_imp}(g), \ref{fig_sqw_kmap_imp}(m), and \ref{fig_sqw_kmap_imp}(s) are common to those in Fig.~\ref{fig_sqw_kmap_random}.
Figures~\ref{fig_sqw_kmap_imp}(a)--\ref{fig_sqw_kmap_imp}(f) present the $\rho$ dependence of ${\cal S}(\bm{q},\omega)$ at $T/J=0.0075$.
We find that the zero-energy peak found in Figs.~\ref{fig_sqw_zero}(e)--\ref{fig_sqw_zero}(h) is almost $\bm{q}$ independent and sharper than that appearing for the strong bond randomness.
This feature survives even at high $T$ as shown in Fig.~\ref{fig_sqw_kmap_imp}.
In addition, there appears another peak at $\omega/J=0.5$, clearly seen for $\rho \gtrsim 0.1$ [Figs.~\ref{fig_sqw_kmap_imp}(d)--\ref{fig_sqw_kmap_imp}(f)].
This dispersionless excitation comes from the local excitation in the isolated dimers discussed above (see also Appendix~\ref{app:detail-Sw}).
Except for the $\bm{q}$ independent excitations at $\omega=0$ and $0.5$, the $T$ evolution of ${\cal S}(\bm{q},\omega)$ for $\rho$ is qualitatively similar to that for the bond randomness.

\subsection{Magnetic susceptibility}
\label{sec:result-suscep}

\begin{figure}[t]
  \begin{center}
  \includegraphics[width=\columnwidth,clip]{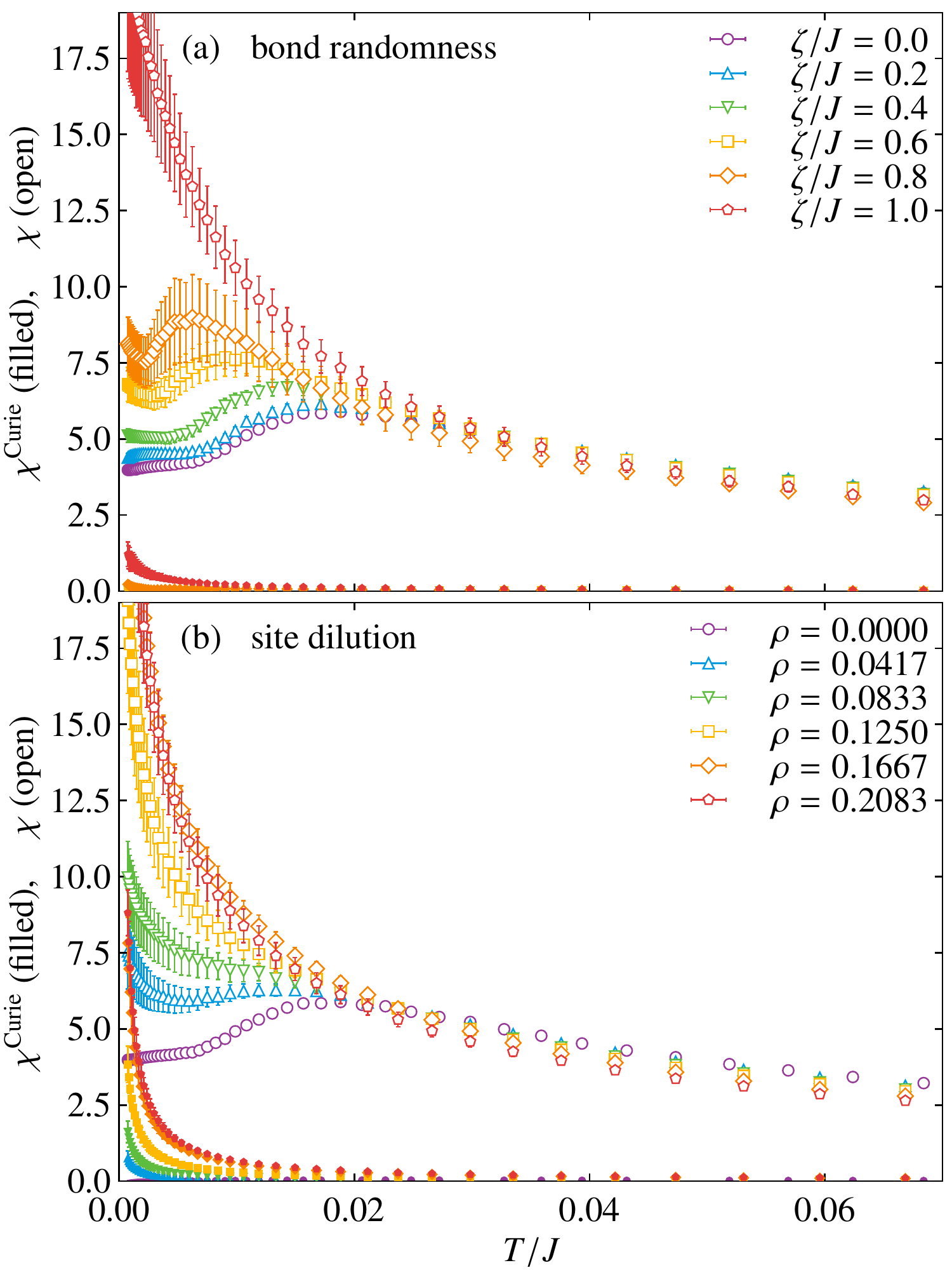}
  \caption{
  $T$ dependences of the magnetic susceptibility $\chi$ (open symbols) and its Curie contribution $\chi^{\rm Curie}$ (filled symbols) for the system with (a) the bond randomness and (b) the site dilution.
}
\label{fig_suscep_linear}
\end{center}
\end{figure}

\begin{figure}[t]
  \begin{center}
  \includegraphics[width=\columnwidth,clip]{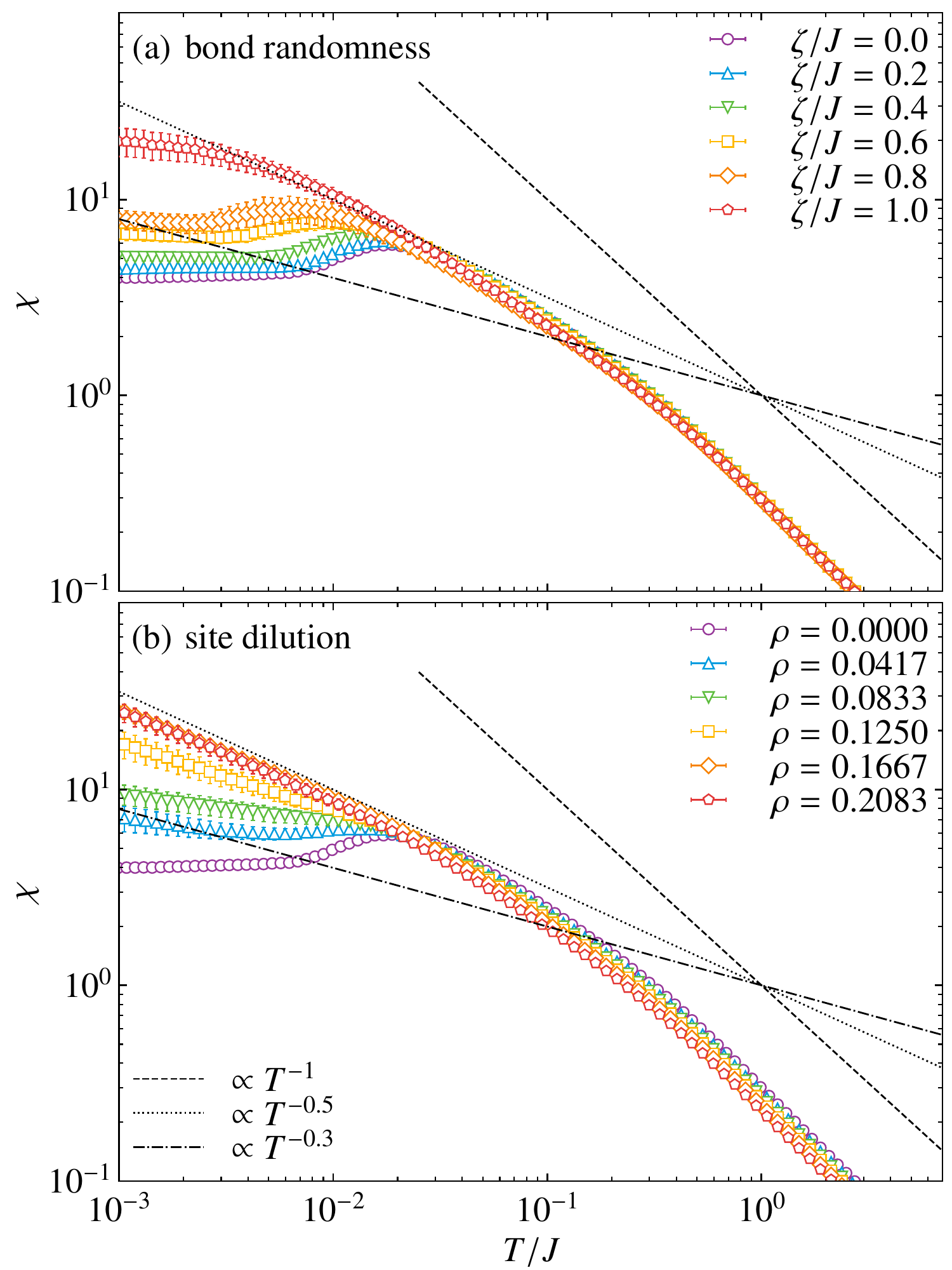}
  \caption{
  Log-log plots of the magnetic susceptibility $\chi$ for the system with (a) the bond randomness and (b) the site dilution.
  The dashed, dotted, and dash-dotted lines represent $\propto 1/T$, $1/T^{0.5}$, and $1/T^{0.3}$, respectively, as the guides for eyes.
  }
  \label{fig_suscep}
  \end{center}
\end{figure}

Next, we examine the disorder effects on the magnetic susceptibility.
In Fig.~\ref{fig_suscep_linear}, we show the results of the isothermal magnetic susceptibility given in Eq.~\eqref{eq:susceptibility}.
In the pristine Kitaev model, while decreasing $T$, the susceptibility increases with obeying the Curie-Weiss law as $1/(4T-J)$, but it deviates from it below $T\simeq T_H$ and eventually converges to a nonzero value similar to the Van Vleck paramagnetism after showing
a peak at $T/J\simeq 0.02$~\cite{yoshitake2016,Yoshitake2017PRBa,Yoshitake2017PRBb}.
The reduction below the peak corresponds to the formation of the flux gap at $T\simeq T_L$.
The effect of the bond randomness is presented in Fig.~\ref{fig_suscep_linear}(a).
While $\chi$ is almost independent of $\zeta/J$ above the peak, the lower-$T$ behavior is highly sensitive to the bond randomness; while increasing $\zeta/J$, the peak of $\chi$ is enhanced and shifted to the lower-$T$ side.
In the low-$T$ region, $\chi$ appears to converge onto a constant for small $\zeta/J \lesssim 0.4$ but it turns to show divergent behavior for larger $\zeta/J$ in the $T$ range calculated here. 
The crossover behavior appears to correlate with the gap closing of the low-energy peak in the dynamical spin structure factor found in Fig.~\ref{fig_sqw_zero}(a) and Figs.~\ref{fig_sqw_kmap_random}(a)--\ref{fig_sqw_kmap_random}(f) as well as the collapse of the half-quantized plateau in the thermal Hall conductivity mentioned in Secs.~\ref{sec:thermodynamics-transport} and \ref{sec:dsf}~\cite{Nasu2020disorder}.

In addition to the isothermal susceptibility, we plot the Curie contribution of the susceptibility, $\chi^{\rm Curie}$ in Eq.~\eqref{eq:suscepcurie}, in Fig.~\ref{fig_suscep_linear}(a).
We find that $\chi^{\rm Curie}$ is much smaller than the rest, that is, the adiabatic susceptibility $\chi^{\rm reg}$ given in Eq.~\eqref{eq:suscepadiabatic}.
The small Curie contribution suggests that the divergent behavior of the low-$T$ $\chi$ for large $\zeta/J$ is not governed by the Curie law proportional to $1/T$.
To see this more clearly, we show the log-log plot of $\chi$ as a function of $T$ in Fig.~\ref{fig_suscep}(a).
No data obey the $1/T$ behavior, including the divergent ones for large $\zeta/J$. 
We note that, at $\zeta/J=1.0$, $\chi$ appears to be proportional to $T^{-0.5}$ in the intermediate-$T$ range at $T/J\simeq0.01$, but a further decrease of $T$ suppresses $\chi$.
However, since the Curie contribution is small but present, we expect that decrease of $T$ renders this dominant.

The effect of the site dilution is distinctly different from that of the bond randomness.
Figure~\ref{fig_suscep_linear}(b) shows the $T$ dependence of $\chi$ in the presence of the site dilution.
While $\chi$ does not strongly depend on $\rho$ above the peak in the pristine case similar to the case of the bond randomness, it shows divergent behavior even for the smallest $\rho$ introduced here, in contrast to the crossover yielded by the bond randomness.
In addition, $\chi^{\rm Curie}$ gives a considerable contribution at low $T$ in comparison with the case of the bond randomness.
Nevertheless, the contribution of $\chi^{\rm reg}$ is still large, and $\chi$ does not show the $T^{-1}$-type divergence in the calculated $T$ range, as shown in the log-log plot in Fig.~\ref{fig_suscep}(b).
This appears to be a power-law divergence although it is hard to precisely determine the asymptotic form in the present numerical analysis.
In the previous studies~\cite{Willans2010,Willans2011}, the logarithmic divergence of $\chi$ was predicted in the compensated case with fixed flux configurations.
The difference might be due to the thermally excited fluxes and/or uncompensated vacancy configurations.
At all events, similar to the bond randomness, the Curie contribution $T^{-1}$ is expected to become dominant eventually in the low-$T$ limit.

\subsection{NMR relaxation rate}
\label{sec:NMR_result}

\begin{figure}[t]
\begin{center}
\includegraphics[width=\columnwidth,clip]{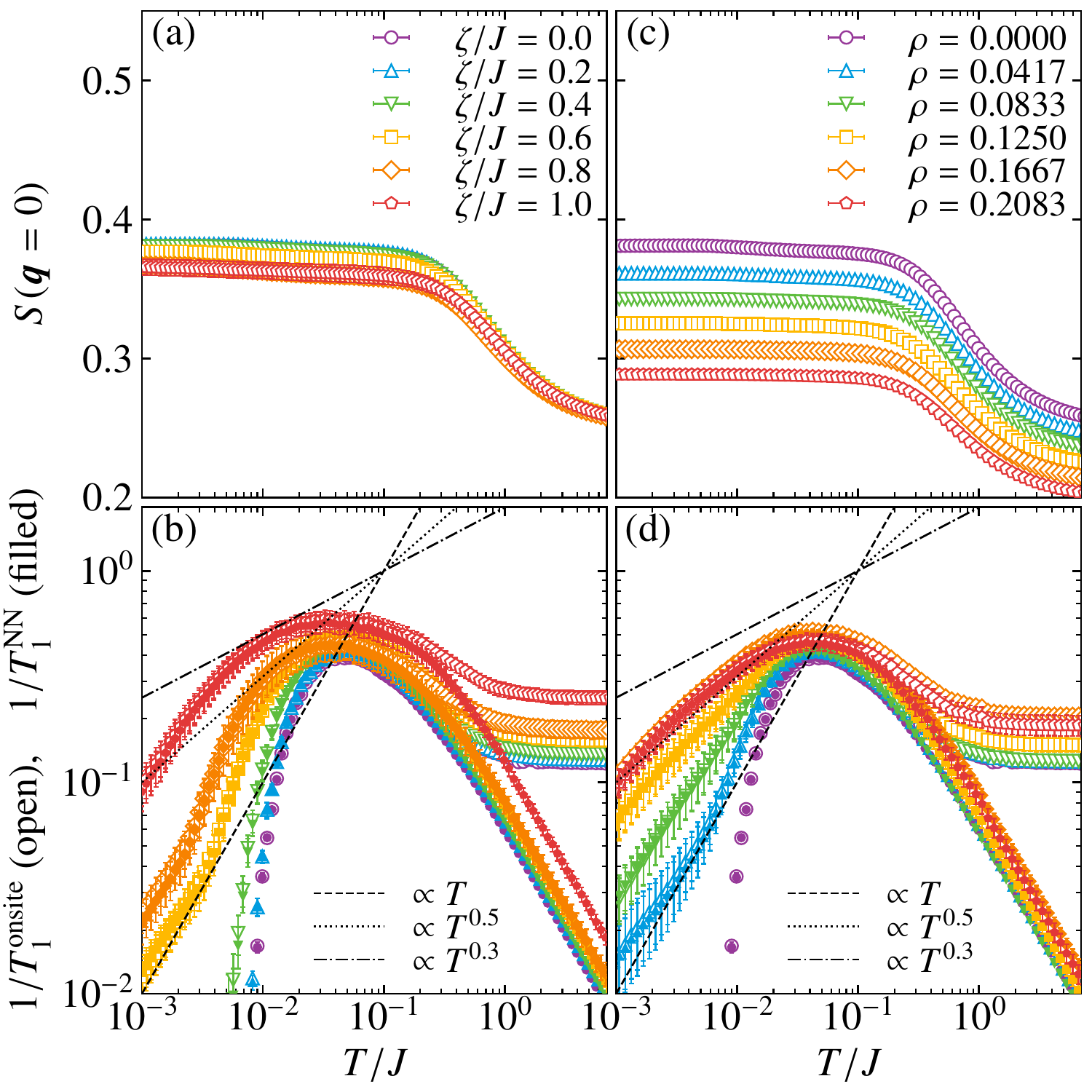}
\caption{
$T$ dependences of (a) the equal-time spin structure factor at $\bm{q}=0$ and (b) the NMR relaxation rates for the onsite and NN sites, $1/T_1^{\rm onsite}$ and $1/T_1^{\rm NN}$, which are represented by the open and filled symbols, respectively, for the systems with the bond randomness. 
(c),(d) Corresponding results for the systems with the site dilution.
The dashed, dotted, and dash-dotted lines in (c) and (d) represent $\propto 1/T$, $1/T^{0.5}$, and $1/T^{0.3}$, respectively, as the guides for eyes.
}
\label{fig_corr}
\end{center}
\end{figure}

Finally, we show the results for the NMR relaxation rate $1/T_1$.
This quantity corresponds to the $\omega\to 0$ limit of the dynamical spin correlations as presented in Eqs.~\eqref{eq:NMRrate_onsite} and \eqref{eq:NMRrate_NN}, and hence, it includes the dynamical nature of the spin correlations.
On the other hand, the static nature is measured by the spin structure factor $S(\bm{q})$ given by Eq.~\eqref{eq:Sq}.
In the pristine Kitaev model, it was found that $1/T_1$ behaves very differently from the uniform component of $S(\bm{q})$; while $S(\bm{q}=0)$ increases from high $T$ and almost saturates to the value at $T=0$ below $T\simeq T_H$~\cite{PhysRevB.92.115122}, $1/T_1$ shows a significant $T$ dependence with showing a peak between $T_H$ and $T_L$ and an exponential decay below $T_L$ due to the flux gap opening~\cite{yoshitake2016,Yoshitake2017PRBa,Yoshitake2017PRBb}. 
These are reproduced in our results for $\zeta/J=0$ and $\rho=0$ shown in Fig.~\ref{fig_corr}.

By introducing the bond randomness, the saturation value of $S(\bm{q}=0)$ is slightly suppressed, while the high-$T$ behavior is almost unchanged, as shown in Fig.~\ref{fig_corr}(a).
In contrast, in the case of the site dilution, $S(\bm{q}=0)$ is largely suppressed by disorder in the whole $T$ region, as shown in Fig.~\ref{fig_corr}(b).
This is because the spin correlations including the vacancy sites vanish, and hence, the averaged correlations decrease with increasing the number of vacancies.

Although the overall $T$ dependence of $S(\bm{q}=0)$ is not altered by the two types of disorder, we find that $1/T_1$ is sensitively influenced by them in a different way, especially at low $T$.
Figure~\ref{fig_corr}(b) shows the $T$ dependences of $1/T_1^{\rm onsite}$ and $1/T_1^{\rm NN}$ given in Eqs.~\eqref{eq:NMRrate_onsite} and \eqref{eq:NMRrate_NN}, respectively, for the case of the bond randomness.
For all $\zeta/J$, the onsite and NN data are overlapped with each other, except for $T\gtrsim T_H$ where $1/T_1^{\rm onsite}$ is almost $T$ independent while $1/T_1^{\rm NN}$ decreases with increasing $T$.  
The high-$T$ behaviors are not altered by the introduction of the bond randomness, while the value increases with $\zeta/J$.
The increases are due to the enhancement of the quasielastic component of ${\cal S}(\bm{q},\omega)$ (Fig.~\ref{fig_sqw_kmap_random}).
On the other hand, although the exponential decay at low $T$ appears to be intact for $\zeta/J\lesssim 0.4$, it is qualitatively changed to a power-law type asymptotic behavior for larger $\zeta/J$; it appears to be fitted by a $T$-linear function, as shown in Fig.~\ref{fig_corr}(b).
The change of the low-$T$ asymptotic behavior is well correlated with the crossover found for $\chi$ [see Fig.~\ref{fig_suscep_linear}(a)].
Because the static spin correlation does not show a drastic change as shown in Fig.~\ref{fig_corr}(a), our results indicate that the crossover is of dynamical nature, closely related with the closing of the spin gap found in Sec.~\ref{sec:dsf}.

In contrast, the introduction of the site dilution appears to immediately change the low-$T$ exponential decay into a power-law one in the calculated parameter range, as shown in Fig.~\ref{fig_corr}(d).
We find that the low-$T$ data is well fitted by a $T$-linear function similar to the case of the strong bond randomness, but the power appears to gradually decrease while increasing $\rho$; the data at $\rho=0.2083$ is fitted by $T^{0.5}$. 
To discuss the low-$T$ behavior more clearly, we need lower-$T$ data with higher accuracy.

\section{Discussion}
\label{sec:discussion}

We compare the results obtained in the present study with the experimental ones for the Kitaev candidate materials. 
As stated in Sec.~\ref{sec:Introduction}, the bond randomness is expected to be introduced in (Na$_{1-x}$Li$_x$)$_2$IrO$_3$~\cite{Cao2013,Manni2014,Rolfs2015,gupta2016raman,Hermann2017,Simutis2018} and H$_3$LiIr$_2$O$_6$~\cite{Kitagawa2018nature}.
While the former shows a magnetic order at a suppressed transition temperature in the intermediate $x$ region, the latter does not exhibit a magnetic order down to the lowest $T$, suggesting the realization of the QSL ground state.
In this material, $1/(T_1 T)$ as well as the Knight shift was reported to be constant at low $T$ in a weak magnetic field~\cite{Kitagawa2018nature}.
In the present calculations, we find that, in the large $\zeta/J$ region, $1/T_1$ appears to be proportional to $T$ as discussed in Sec.~\ref{sec:NMR_result}, meaning that $1/(T_1 T)$ is almost constant for strong bond randomness.
This appears to be consistent with the experimental result.
However, the present calculations also indicate that the magnetic susceptibility is divergent in this region as discussed in Sec.~\ref{sec:result-suscep}, which is incompatible with the experimental result of the Knight shift.
While this might be due to the subsidiary non-Kitaev interactions, further researches are needed to clarify the discrepancy between experiment and theory.
Meanwhile, the systematic NMR study has not been performed for (Na$_{1-x}$Li$_x$)$_2$IrO$_3$ thus far to the best of our knowledge. 
The detailed comparison with the magnetic susceptibility would be helpful to discuss the effect of bond randomness.

On the other hand, the effect of the site dilution is experimentally introduced by the replacement of the magnetic ions by nonmagnetic ones, e.g., in $A_2$LiO$_3$ by replacing Ir$^{4+}$ to Ru$^{4+}$~\cite{Lei2014} and Ti$^{4+}$~\cite{Manni2014a}, and in $\alpha$-RuCl$_3$ by replacing Ru$^{3+}$ to nonmagnetic Ir$^{3+}$~\cite{Lampen-Kelley2017,Do2018,Do2020}. 
While the former shows a spin glass or dimerized behavior, the latter realizes a nonmagnetic state down to the lowest $T$, where the magnetic susceptibility shows power-law divergence proportional to $\sim T^{-0.25}$~\cite{Do2018,Do2020}.
This behavior might be related to the present result with a power-law divergence shown in Fig.~\ref{fig_suscep}(b).
Also for this case, however, further systematic study together with the NMR measurement is desirable to clarify the effect of site dilution.

Our results for the site dilution suggest that the low-$T$ measurements of the spin dynamics would be a good probe of the Majorana zero modes induced by the site vacancies. 
This was pointed out for $\chi$ in the case of a single vacancy~\cite{Willans2010,Willans2011,Santhosh2012}, but our study clarify that the site dilution with nonzero density leads to the characteristic $T$ dependences in the dynamical spin structure factor and $1/T_1$ as well as $\chi$, while the existence of the Majorana zero modes in the thermodynamic limit is still in debate as mentioned in Sec.~\ref{sec:dsf}. 
In particular, the power-law like behaviors in $\chi$ and $1/T_1$ are worth investigating, by carefully controlling the strength of disorder. 
We note, however, that our results are for the model with the Kitaev interactions only; further study by including non-Kitaev interactions is necessary for the detailed comparison with experiments.

\section{Summary}
\label{sec:summary}

In summary, we have studied the spin dynamics of the Kitaev model in the presence of the two types of disorder, bond randomness and site dilution, using the quantum Monte Carlo simulations.
For the bond randomness, we found that the low-energy peak in the dynamical spin structure factor is shifted to the lower-energy side, and the spin gap appears to be closed, while the high-energy continuum does not change significantly. 
The results suggest that the bond randomness affects the flux excitations dominantly rather than the Majorana fermion excitations.
On the other hand, the site dilution suppresses the low-energy peak without noticeable energy shift, but instead, develops additional lower-energy peaks including the zero-energy one originating from the Majorana zero modes induced around the vacancies.
The distinctly different behaviors in the low-energy spin excitations also manifest in the characteristic temperature dependences of the magnetic susceptibility and the NMR relaxation rate.
We show that the low-temperature magnetic susceptibly undergoes a crossover from convergent to divergent behavior with increasing the disorder in the calculated temperature range.
Similar crossover is also observed in the NMR relaxation rate from exponential decay to power-law one.
Since the static spin correlations are almost unchanged by the disorder, the results indicate that the crossover originates from the dynamical nature associated with the closing of the spin gap.
In the case of the site dilution, the disorder effect immediately appears in the magnetic susceptibility and the NMR relaxation rate, in contrast to the bond randomness, presumably because of the appearance of the Majorana zero modes.
The present results unveiled the qualitative difference between the two types of disorder in the spin dynamics and will stimulate further experimental studies on the disorder effects on the Kitaev magnets.
They will also provide a solid ground for discussing the effect of non-Kitaev interactions in the disordered Kitaev systems. 

\begin{acknowledgments}
The authors thank J.~Knolle, R.~Moessner, and K.~Nomura for fruitful discussions. 
Parts of the numerical calculations were performed in the supercomputing
systems in ISSP, the University of Tokyo.
This work was supported by Grant-in-Aid for Scientific Research from
JSPS, KAKENHI Grant Nos.~JP16H02206, JP18H04223, JP19K03742, JP19H05825, and JP20H00122, and by JST PREST (JPMJPR19L5) and JST CREST (JPMJCR18T2).
\end{acknowledgments}

\appendix

\section{Dynamical spin correlation at $T=0$}
\label{app:zeroT}

In this appendix, we derive the $T\to 0$ forms of the dynamical spin correlations in Eqs~\eqref{eq:corr_orig1} and \eqref{eq:corr_orig2}.
For preparation, we note that Eq.~\eqref{eq:diagiA} includes the positive eigenvalues only, but the complete set of the eigenvalues of the $N_{\rm spin}\times N_{\rm spin}$ Hermitian matrix $iA$ appear in a pairwise fashion as 
\begin{align}
  {\cal U}^\dagger iA {\cal U}={\rm diag}\{\varepsilon_1,\varepsilon_2,\cdots,\varepsilon_{N_{\rm spin}/2}, -\varepsilon_1,-\varepsilon_2,\cdots,-\varepsilon_{N_{\rm spin}/2}\},
\end{align}
where ${\cal U}$ is a $N_{\rm spin}\times N_{\rm spin}$ unitary matrix given by using $U$ in Eq.~\eqref{eq:diagiA} as
\begin{align}
  {\cal U}=\begin{pmatrix}
    U& U^*
  \end{pmatrix}.
\end{align}
Note that $U^\dagger U=\mathbb{1}$ and $U^T U=\mathbb{0}$ as ${\cal U}$ is unitary, where $\mathbb{1}$ and $\mathbb{0}$ are the $N_{\rm spin}/2\times N_{\rm spin}/2$ unit and zero matrices, respectively.

For the derivation of the dynamical spin correlations in the $T\to 0$ limit, we start with rewriting
the $N_{\rm spin}\times N_{\rm spin}$ matrices, $X$ and $Y(t)$ in Eqs.~\eqref{eq:Xform} and \eqref{eq:Yform}, respectively, into 
\begin{align}
  X&=UU^\dagger, \\ 
  Y(t)&=
  \left(U^{(r)} e^{-it \Lambda^{(r)}} U^{(r)\dagger} 
  +
  U^{(r)*} e^{it \Lambda^{(r)}} U^{(r)T} 
  \right),
\end{align}
and similarly, $C'(t)$ in Eq.~\eqref{eq:Cp} into
\begin{align}
  C'(t)=U^*e^{-it \Lambda}U^T.
\end{align}
Here, we introduce the $N_{\rm spin}/2\times N_{\rm spin}/2$ diagonal matrices $\Lambda$ and $\Lambda^{(r)}$ as
\begin{align}
  \Lambda&={\rm diag}\{\varepsilon_1,\varepsilon_2,\cdots,\varepsilon_{N_{\rm spin}/2}\},
\\
  \Lambda^{(r)}&={\rm diag}\{\varepsilon_1^{(r)}, \varepsilon_2^{(r)},\cdots,\varepsilon_{N_{\rm spin}/2}^{(r)}\},
\end{align}
respectively.
To calculate $C(t)=X+C'(t)Y(t)$ in the $T\to 0$ limit, we introduce the following four $N_{\rm spin}/2\times N_{\rm spin}/2$ matrices:
\begin{align}
  {\cal M}(t)&=U^T  Y(t)U^*,\quad {\cal M}'(t)=U^T C(t)U^*,\\
  {\cal N}(t)&=U^T  Y(t)U,\quad {\cal N}'(t)=U^T C(t)U.
\end{align}
Using $U^T U=\mathbb{0}$, we obtain the following relations:
\begin{align}
  {\cal M}'(t) &= U^T C'(t) Y(t)U^* = e^{-it \Lambda} {\cal M}(t),\\
  {\cal N}'(t)&=e^{-it \Lambda} {\cal N}(t).
\end{align}
We also find that $U^\dagger C(t)U^*=\mathbb{0}$ and $U^\dagger C(t)U=\mathbb{1}$.
From the above relations, we obtain
\begin{align}
  {\cal U}^T C(t){\cal U}^* =
  \begin{pmatrix}
    {\cal M}'(t)&{\cal N}'(t)\\ \mathbb{0}&\mathbb{1}
  \end{pmatrix}\equiv F(t).
\label{eq:defF}
\end{align}
This leads to the relation ${\rm det}C(t)={\rm det}{\cal M}'(t)$.

To obtain $C(t)^{-1}$, we use the general formula
\begin{align}
\begin{pmatrix}
  A&B\\C&D
\end{pmatrix}^{-1}
=
\begin{pmatrix}
  A^{-1} + A^{-1}B V^{-1}CA^{-1} & -A^{-1} B V^{-1}\\
  -V^{-1}CA^{-1} &V^{-1}
\end{pmatrix},
\end{align}
which holds when $A$ and $V\equiv D-CA^{-1}B$ are regular matrices.
By applying this formula to $F(t)$ in Eq.~\eqref{eq:defF}, we find 
\begin{align}
  F(t)^{-1}=
  \begin{pmatrix}
    {\cal M}'(t)^{-1}&-{\cal M}'(t)^{-1}{\cal N}'(t)\\
    \mathbb{0}&\mathbb{1}
  \end{pmatrix}
  ={\cal U}^T C^{-1}(t){\cal U}^*.
\end{align}
Using this result, we find
\begin{align}
  C(t)^{-1}C'(t) = U^* {\cal M}(t)^{-1}U^T
.
\end{align}
This leads us to obtain
\begin{align}
  \sqrt{{\rm det} C(t)}C(t)^{-1} C'(t)
  &=\sqrt{{\rm det}e^{-i \Lambda t}{\rm det} {\cal M}(t)}U^*{\cal M}(t)^{-1} U^T\nonumber\\
  &=e^{iE_0 t}\sqrt{{\rm det} {\cal M}(t)}U^*{\cal M}(t)^{-1} U^T,
\end{align}
where $E_0$ is the ground state energy given by
\begin{align}
E_0=-\frac{1}{2}\sum_{\lambda=1}^{N_{\rm spin}/2}\varepsilon_\lambda.
\end{align}
Therefore, the dynamical correlation functions in the $T\to 0$ limit are given by 
\begin{align}
  \means{S_j^z(t) S_{j}^z}&=\frac{1}{2}e^{iE_0 t}\sqrt{{\rm det} {\cal M}(t)}\left[U^*{\cal M}(t)^{-1} U^T\right]_{jj},\label{eq:corr1}\\
  \means{S_j^z(t) S_{j'}^z}&=\frac{i\eta_r}{2}e^{iE_0 t}\sqrt{{\rm det} {\cal M}(t)}\left[U^*{\cal M}(t)^{-1} U^T\right]_{j'j}\nonumber\\
  &=-\frac{i\eta_r}{2}e^{iE_0 t}\sqrt{{\rm det} {\cal M}(t)}\left[U^*{\cal M}(t)^{-1} U^T\right]_{jj'},\label{eq:corr2}
\end{align}
where $j$ and $j'$ in Eq.~\eqref{eq:corr2} are the $A$ and $B$ sublattice sites on the $z$ bond $r$, respectively.
We note that by introducing the $N_{\rm spin}/2\times N_{\rm spin}/2$ matrices, ${\cal X}$ and ${\cal Y}$, as
\begin{align}
  {\cal X}=U^{(r)T}U^*,\qquad 
  {\cal Y}=U^{(r)T}U,
\end{align}
respectively, ${\cal M}(t)$ can be written as
\begin{align}
  {\cal M}(t)&={\cal Y}^T e^{-it \Lambda^{(r)}} {\cal Y}^*
  +{\cal X}^\dagger e^{it \Lambda^{(r)}} {\cal X},
\end{align}
which corresponds to the form given in Refs.~\cite{PhysRevLett.112.207203,PhysRevB.92.115127}.

\section{Details of dynamical spin correlations}
\label{app:detail-Sw}

\begin{figure}[t]
  \begin{center}
  \includegraphics[width=\columnwidth,clip]{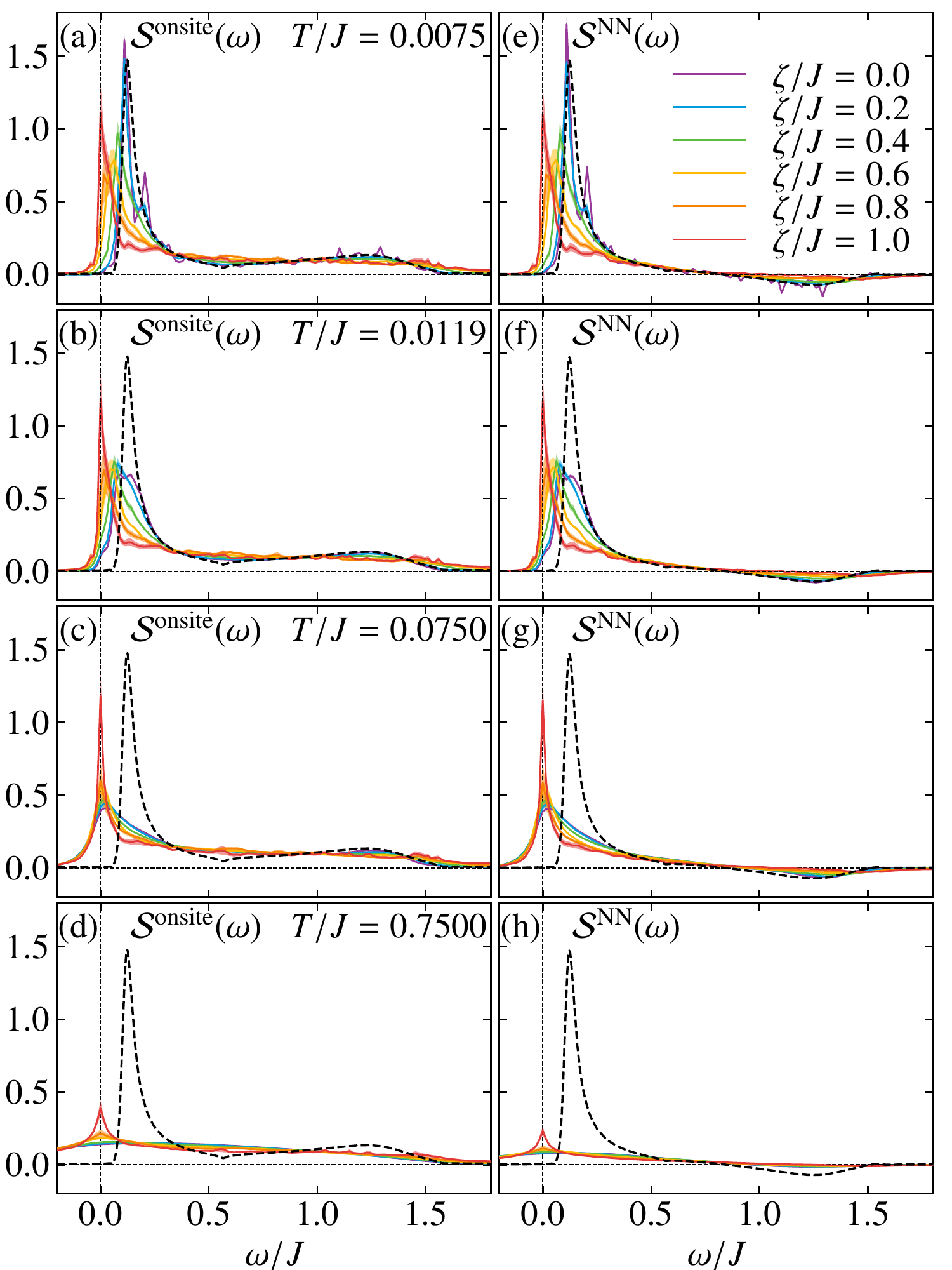}
  \caption{
    (a)--(d) Frequency dependences of the onsite dynamical spin correlation ${\cal S}^{\rm onsite}(\omega)$ for the disordered Kitaev model while changing the bond randomness $\zeta/J$ at (a) $T/J=0.0075$, (b) $T/J=0.0119$, (c) $T/J=0.0750$, and (d) $T/J=0.7500$.
    (e)--(h) Corresponding results for the NN component ${\cal S}^{\rm NN}(\omega)$.
    The notations are common to those in Fig.~\ref{fig_sqw_zero}.
  }
  \label{fig_s_random}
  \end{center}
  \end{figure}
  
  \begin{figure}[t]
  \begin{center}
  \includegraphics[width=\columnwidth,clip]{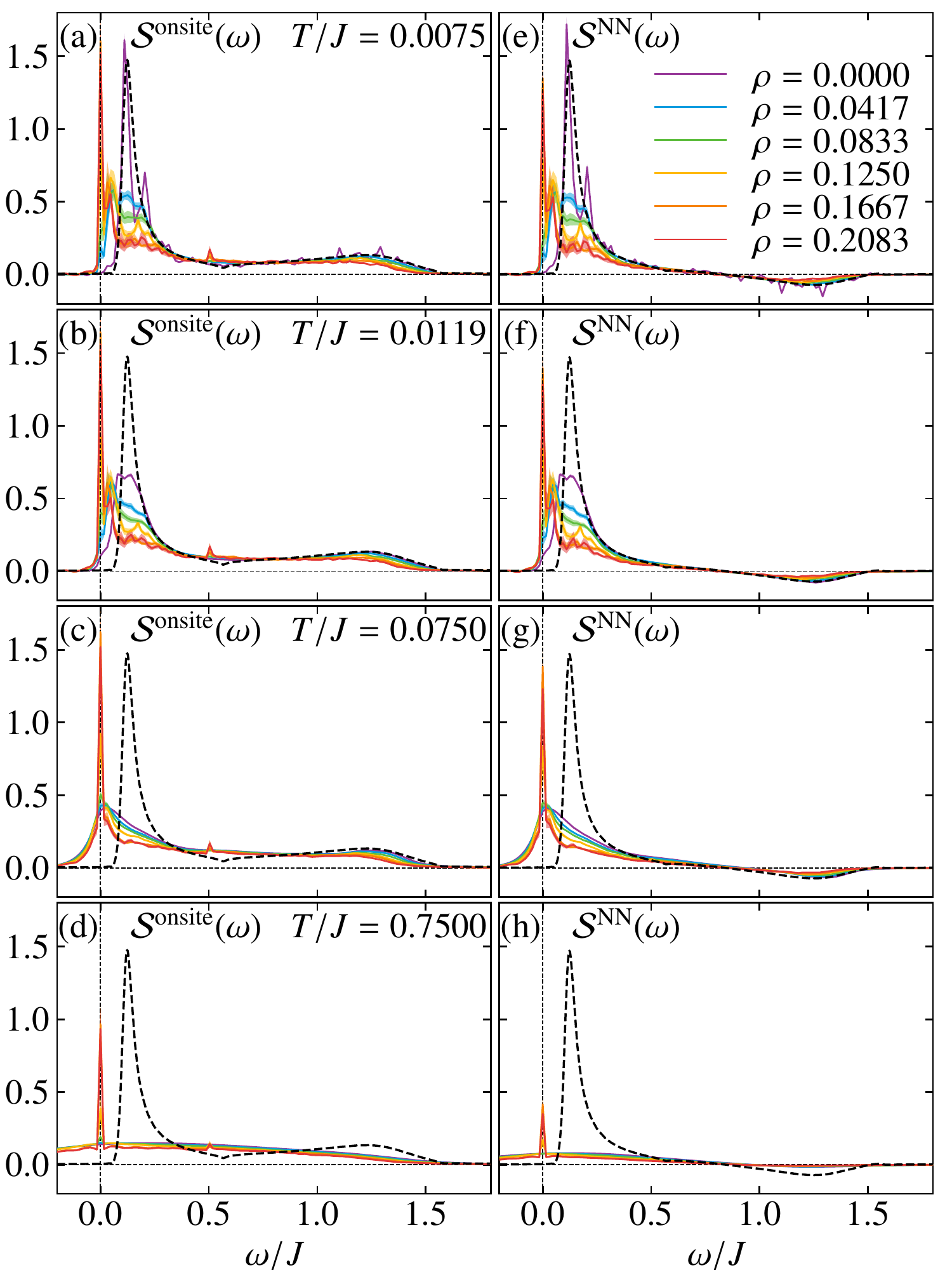}
  \caption{
Similar plots to Fig.~\ref{fig_s_random} for the case of the site dilution.
  }
  \label{fig_s_imp}
  \end{center}
  \end{figure}

  \begin{figure}[t]
  \begin{center}
  \includegraphics[width=\columnwidth,clip]{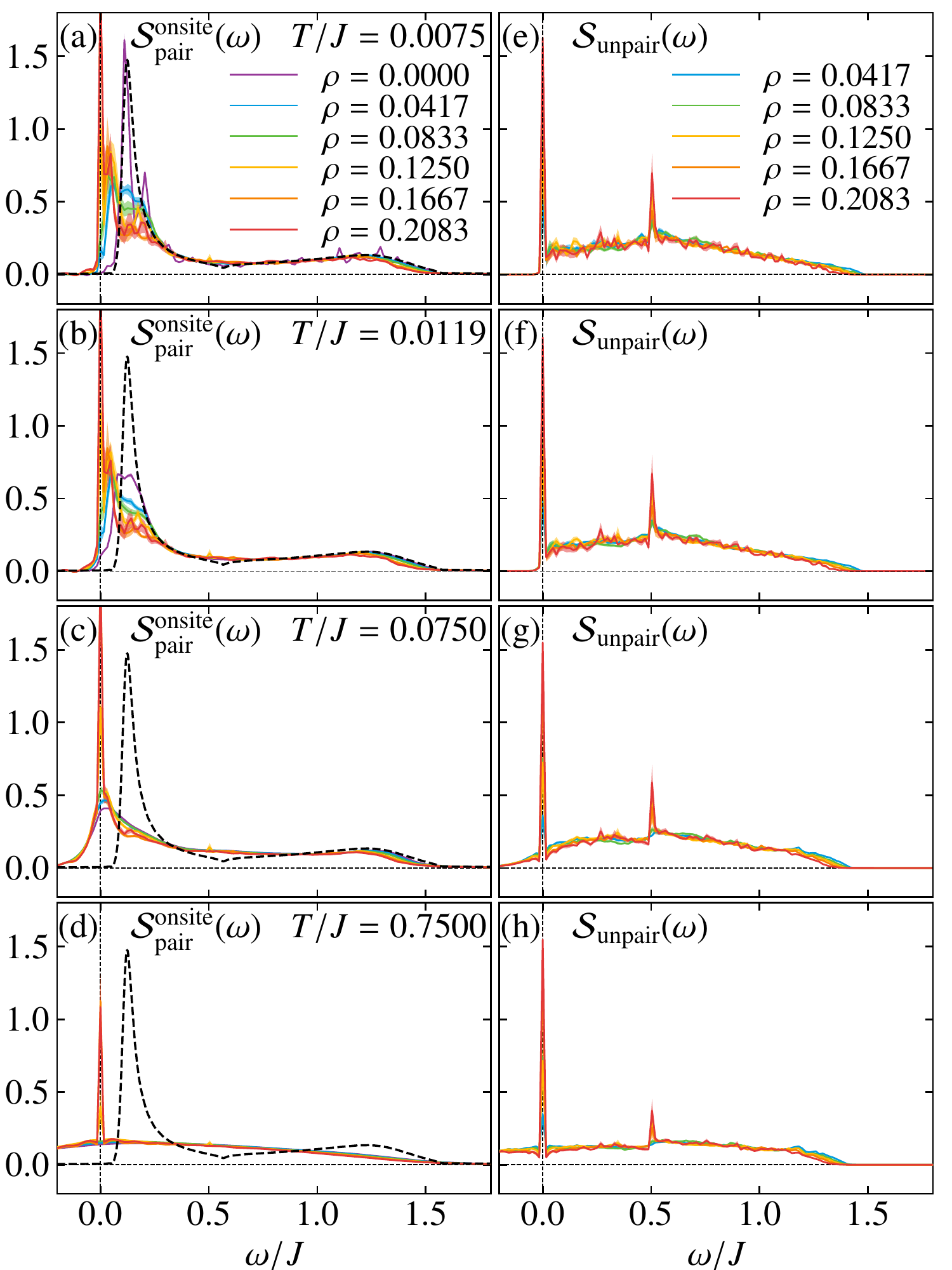}
  \caption{
Decomposition of ${\cal S}^{\rm onsite}(\omega)$ in Fig.~\ref{fig_s_imp} into the contributions of (a)--(d) ${\cal S}_{\rm pair}^{\rm onsite}(\omega)$ and (e)--(h) ${\cal S}_{\rm unpair}(\omega)$.
  }
  \label{fig_s_imp-each}
  \end{center}
  \end{figure}

In this appendix, we show the quantum Monte Carlo results for the dynamical spin correlations decomposed into the onsite and NN components in Eqs.~\eqref{eq:SLOw} and \eqref{eq:SNNw}, respectively. 
The results are shown in Figs.~\ref{fig_s_random} and \ref{fig_s_imp} for the cases of the bond randomness and site dilution, respectively.
For the low-energy part, the two components show similar $T$ and disorder dependences, which are reflected in the dynamical spin structure factor discussed in Sec.~\ref{sec:dsf}.
On the other hand, for the high-energy structure, the difference between ${\cal S}^{\rm onsite}(\omega)$ and ${\cal S}^{\rm NN}(\omega)$ becomes more apparent at higher $T$; the high-energy intensity in ${\cal S}^{\rm NN}(\omega)$ becomes almost zero above $T>T_H$, while the broad spectrum remains almost intact in ${\cal S}^{\rm onsite}(\omega)$.
The contrasting behaviors result in the disappearance of the hour-glass-like continuum in ${\cal S}(\bm{q},\omega)$.
In the case of the site dilution, while the NN component ${\cal S}^{\rm NN}(\omega)$ is proportional to ${\cal S}_{\rm pair}^{\rm NN}(\omega)$ defined in Eq.~\eqref{eq:Snnw}, the onsite one ${\cal S}^{\rm onsite}(\omega)$ can be further decomposed into the two contributions, ${\cal S}_{\rm pair}^{\rm onsite}(\omega)$ and ${\cal S}_{\rm unpair}(\omega)$ in Eqs.~\eqref{eq:Slocw} and \eqref{eq:Sunpw}, respectively.
Figures~\ref{fig_s_imp-each}(a)--\ref{fig_s_imp-each}(d) show the $T$ evolution of ${\cal S}_{\rm pair}^{\rm onsite}(\omega)$.
The results look similar to those in Figs.~\ref{fig_s_imp}(a)--\ref{fig_s_imp}(d) because ${\cal S}_{\rm pair}^{\rm onsite}(\omega)$ gives a dominant contribution to ${\cal S}^{\rm onsite}(\omega)$.
Meanwhile, as shown in Figs~\ref{fig_s_imp-each}(e)--\ref{fig_s_imp-each}(h), ${\cal S}_{\rm unpair}(\omega)$ exhibits a peak at $\omega=0.5$ in addition to the zero-energy one.
The contribution from ${\cal S}_{\rm unpair}(\omega)$ increases with increasing $\rho$, and thereby, the peak at $\omega=0.5$ is more clearly observed in ${\cal S}^{\rm onsite}(\omega)$ for larger $\rho$.
Since the excitation energy is $J/2$, this peak is deduced to originate from the excitations in isolated dimers surrounded by four vacancies.

\section{Effect of configuration of vacancies}
\label{app:compensate}

\begin{figure*}[t]
  \begin{center}
  \includegraphics[width=2\columnwidth,clip]{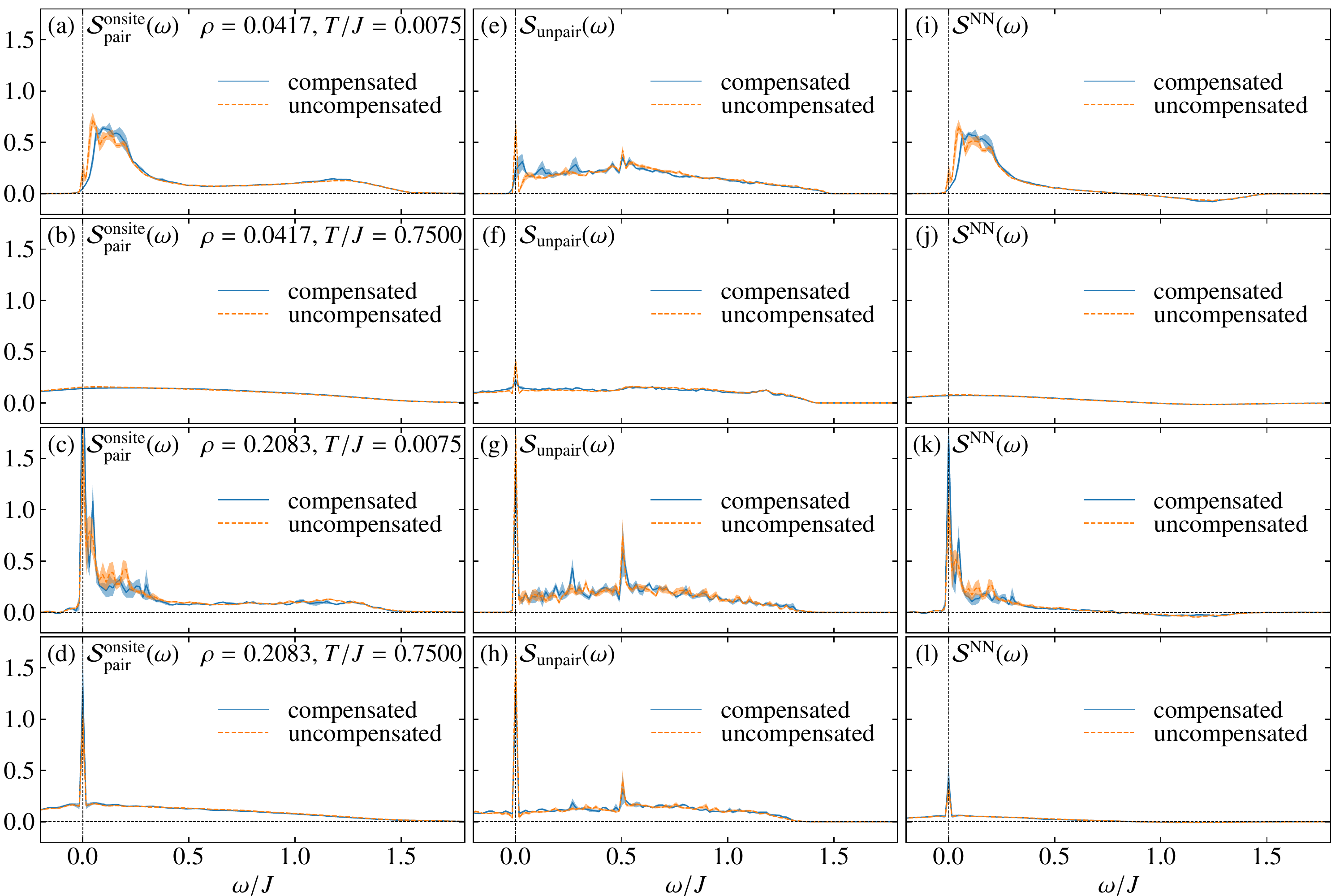}
  \caption{
    (a)--(d) Frequency dependences of the onsite dynamical spin correlation ${\cal S}_{\rm pair}^{\rm onsite}(\omega)$ at (a) $(\rho,T/J)=(0.0417,0.0075)$, (b) $(0.0417,0.7500)$  (c) $(0.2083,0.0075)$, and (d) $(0.2083,0.7500)$ for the compensated and uncompensated vacancy configurations in the systems with site dilution.
    (e)--(h) and (i)-(l) Corresponding results for ${\cal S}_{\rm unpair}(\omega)$ and ${\cal S}^{\rm NN}(\omega)$, respectively.
The notations are common to those in Fig.~\ref{fig_sqw_zero}.
} 
  \label{fig_comp}
  \end{center}
\end{figure*}

In this appendix, we present the analysis of the zero-energy modes appearing in the case of the site dilution. 
As discussed in Secs.~\ref{sec:MC-simulations} and \ref{sec:dsf}, 
the zero-energy modes appear in the uncompensated cases with $N_A\neq N_B$. 
To examine the effect of the zero-energy modes on the dynamical spin correlations, 
we evaluate ${\cal S}_{\rm pair}^{\rm onsite}(\omega)$, ${\cal S}^{\rm NN}(\omega)$, and ${\cal S}_{\rm unpair}(\omega)$ by averaging the contributions for the compensated ($N_A=N_B$) and uncompensated ($N_A\neq N_B$) cases separately among the 10 configurations of vacancies prepared in the calculations.
Figure~\ref{fig_comp} shows the results.
While the data for the two cases almost coincide with each other for relatively large $\rho=0.2083$ [Figs.~\ref{fig_comp}(c), \ref{fig_comp}(d) \ref{fig_comp}(g) \ref{fig_comp}(h) \ref{fig_comp}(k), and \ref{fig_comp}(l)], notable differences are observed for small $\rho$ and low $T$, as shown in Figs.~\ref{fig_comp}(a), \ref{fig_comp}(e), and \ref{fig_comp}(i).
In ${\cal S}_{\rm pair}^{\rm onsite}(\omega)$ and ${\cal S}^{\rm NN}(\omega)$, the low-energy weight in the uncompensated case is considerably larger than that in the compensated case [Figs.~\ref{fig_comp}(a) and \ref{fig_comp}(i)].
On the other hand, in ${\cal S}_{\rm unpair}(\omega)$, there is a peak at $\omega=0$ due to the zero modes whose number is proportional to $|N_A-N_B|$, and the gap structure is seen above this peak in the uncompensated case at low $T$ [Fig.~\ref{fig_comp}(e)].
These behaviors might correspond to those in the density of states for graphene with the low vacancy density~\cite{Pereira2006,Pereira2008,ShangduanWu2008,Hafner2014}.
While it is difficult to conclude the existence of the zero-energy peak in our calculations on the finite-size cluster, the zero-energy peak is expected to remain even in the thermodynamic limit for sufficiently large $\rho$, because it is commonly present in both compensated and uncompensated cases as demonstrated for $\rho=0.2083$.

\bibliography{refs}

\end{document}